\documentclass[a4paper,14pt]{article}
\usepackage[english]{babel}
\pdfoutput=1 

\usepackage{jheppub} 

\usepackage[T1]{fontenc} 

\usepackage[T1,T2A]{fontenc}

\usepackage{comment}

\usepackage{cases}
\usepackage{amsmath}

\usepackage{stackrel}

\usepackage{tikz}
\usetikzlibrary{matrix,arrows.meta}

\newcommand{\beq}{\begin{equation}}
    \newcommand{\eeq}{\end{equation}}
\newcommand{\bi}{\begin{itemize}}
    \newcommand{\ei}{\end{itemize}}
\newcommand{\bea}{\begin{eqnarray}}
    \newcommand{\eea}{\end{eqnarray}}
\newcommand{\bt}{\begin{tabular}}
    \newcommand{\et}{\end{tabular}}
\newcommand{\bc}{\begin{center}}
    \newcommand{\ec}{\end{center}}

\newcommand{\ket}[1]{|#1\rangle}

\newcommand{\be}{\begin{equation}}
    \newcommand{\ee}{\end{equation}}
\newcommand{\ba}{\begin{array}}
    \newcommand{\ea}{\end{array}}

\newcommand{\lb}[1]{\label{#1}}

\def\bbox{{\,\lower0.9pt\vbox{\hrule \hbox{\vrule height 0.2 cm
                \hskip 0.2 cm \vrule height 0.2 cm}\hrule}\,}}
\newcommand{\dsl}{\pa \kern-0.5em /}

\newcommand{\nn}{\nonumber \\}

\makeatletter \@addtoreset{equation}{section} \makeatother

\def\slashchar#1{\setbox0=\hbox{$#1$}           
    \dimen0=\wd0                                 
    \setbox1=\hbox{/} \dimen1=\wd1               
    \ifdim\dimen0>\dimen1                        
    \rlap{\hbox to \dimen0{\hfil/\hfil}}      
    #1                                        
    \else                                        
    \rlap{\hbox to \dimen1{\hfil$#1$\hfil}}   
    /                                         
    \fi}

\pdfoutput=1

\title{\boldmath Casimir operators of $4D\,, \, \mathcal{N}=2$ supersymmetry  in the harmonic approach}

\author[a,b]{Egor~Eremeev,}
\author[a,b]{Evgeny~Ivanov}

\affiliation[a]{Bogoliubov Laboratory of Theoretical Physics, JINR,\\141980 Dubna, Moscow region, Russia}
\affiliation[b]{Moscow Institute of Physics and Technology,\\ 141700 Dolgoprudny, Moscow region, Russia}

\emailAdd{eremeev.ei@phystech.edu}
\emailAdd{eivanov@theor.jinr.ru}

\abstract{We construct Casimir operators of $4D, \,\mathcal{N}=2$ supersymmetry algebra
in the harmonic superspace approach, and provide the explicit expressions for them in terms of covariant derivatives
in the analytic basis. Specifically, the superisospin operator $C_I$ is expressed in terms of new 'long'
harmonic derivatives $\nabla^{\pm\pm}, \nabla^0$ as  $C_I = \frac{1}{4}\left(\nabla^0\right)^2 + \frac{1}{2}\left\{\nabla^{++},\nabla^{--}\right\}$.
These derivatives involve non-local terms with the inverse Box operator $\Box^{-1}$ and  satisfy modified $SU(2)$ algebra relations
in which the eigenvalues of $\nabla^0$ on analytic superfields are shifted by -2 relative to the standard $U(1)$ harmonic charges.
We also construct $\mathcal{N}=2$ superhelicity and super-isohelicity operators relevant to the massless case (with vanishing $\Box$) and find their
eigenvalues for few instructive examples of $4D, \,\mathcal{N}=2$ superfield theories.}

\makeatletter
\gdef\@fpheader{}
\makeatother

\begin{document}
\maketitle
\flushbottom

\section{Introduction}
Multiplets of $4D, \,{\cal N}=2$ Poincar\'e supersymmetry are characterized by the eigenvalues of the invariant mutually commuting Casimir-type operators. Off shell
and in the massive on-shell case these are the superspin and
superisospin operators \cite{SS,Sokatchev:1975gg,Rittenberg:1981cp}, while in the massless on-shell case the superspin is substituted by the superhelicity operator \cite{Gates,Buchbinder:1998twe}.
Also, there exist some additional Casimir operators related to the extra $U(1)$ R-symmetry of $4, \,{\cal N}=2$ superalgebra. Apart from these operators there is the standard square-mass
operator $\Box \equiv - P^mP_m$ which, in contradistinction to the superspin-square operator,
does not take definite values on the supermultiplets
in the interaction case \footnote{It is similar to the situation in the purely bosonic $4D$ theories where the square-mass operator $P^2$ takes definite values only on the free on-shell fields,
while the square-spin operator
can be defined for any off-shell Lorentz covariant field \cite{OgPol}.}. It is very important to know the explicit expressions of the full set of Casimir operators in the superfield approach.
For instance,  in the quantum
case they are needed for constructing various projection operators  in the superfield propagators (see, e.g., \cite{Buchbinder:1998twe,Ponds,ProjUse}).

At present, the most natural and widely recognized way of dealing with $4D, \,{\cal N}=2$ theories is the Harmonic Superspace (HSS) approach, which makes manifest the underlying
${\cal N}=2$ harmonic analyticity
of these theories \cite{HSS,Book}. Besides the ordinary ${\cal N}=2$ superspace coordinates $(x^{\alpha\dot\alpha}, \theta^{\alpha i}, \bar\theta^{\dot\alpha}_j)$
(here the indices $i, j $ stand for the doublet indices
of the automorphism $R$ symmetry group $SU(2)_{A}$), HSS involves isospinor harmonic coordinates $u^{\pm i}, u^{+i}u_{i}^- =1$, parametrizing the internal 2-sphere $S^2 \sim SU(2)_{A}/U(1)_{A}$.
It is of obvious interest to find the explicit expressions for super Casimir operators acting on the harmonic ${\cal N}=2$ superfields and on their most important subclass consisting of the analytic
harmonic superfields. The latter depend on the half of the original Grassmann  coordinates and are the basic ingredients of all interesting ${\cal N}=2$ theories
(super Yang-Mills, supergravities and  matter hypermultiplets).
The primary incentive of the present  paper is to construct such a formalism and  to highlight its some peculiar features.

While in refs. \cite{Sokatchev:1975gg,Rittenberg:1981cp} the super Casimir operators were defined through the differential operators  realized in the conventional
${\cal N}=2$ superspace (ordinary and spinorial covariant derivatives),
the HSS approach requires additional derivatives, those with respect to the new harmonic variables. These harmonic derivatives will be shown
to play  a crucial  role in the structure of superisospin Casimirs applied to the harmonic superfields.

Though the considerations of the original ref. \cite{Rittenberg:1981cp} are in principle applicable to any $4D, {\cal N}$ Poincar\'e supersymmetries, our study here is limited to the
${\cal N}=2$ case, where the harmonic formalism was worked out in most details. Possible extensions to other kinds of harmonic superspaces, e.g., to $4D, \,{\cal N}=3$ harmonic superspace \cite{N3HSS},
will be studied elsewhere.

The paper is organized as follows. Basic definitions and notions including the basics of the harmonic superspace are the contents of section 2. In section 3 the general definitions
of the superspin, superisospin and $R$-symmetry $U(1)$ Casimir operators, both in the conventional and harmonic ${\cal N}=2$ superspaces, are given and their realization on off-shell harmonic analytic
superfields is presented. It is shown that the superisospin Casimir is naturally expressed in terms of some nonlocal generalization of the harmonic derivatives, $\nabla^{\pm\pm}, \nabla^0$.
They satisfy the same $SU(2)$ algebra as the standard harmonic derivatives $D^{\pm\pm}, D^0$, but with a shift by $-2$ for eigenvalues of $\nabla^0$ on analytic harmonic superfields
compared to those of $D^0$. In section 4 the superisospin decompositions of analytic off-shell superfields are explicitly given, with some particular cases as illustrations. In section 5 we
generalize our consideration to the massless on-shell case, where the previously defined operators $\nabla^{\pm\pm}, \nabla^0$ become inapplicable because of the presence of singular factors $\Box^{-1}$.
These are replaced by the proper ``super-isohelicity'' operators for which we establish the algebra which proves to be  a non-trivial generalization of $su(2)$.  Its invariants, along with the
super-helicity and its $U(1)$ $R$-symmetry analog, fully characterize the on-shell harmonic superfields. In section 6 a few instructive examples of free ${\cal N}=2$ supersymmetric theories
in the harmonic approach are treated from this novel point of view. These are the $q^+$ and $\omega$ hypermultiplets, as well as the abelian ${\cal N}=2$ gauge theory\footnote{Applications for the most
 interesting cases of ${\cal N}=2$ supergravity in the harmonic approach \cite{HSS,Book}, as well as its extension to higher spins \cite{BIZ1,SupergravityMotionEquations}, will be discussed elsewhere.}.
 In Appendices A and B
we present some useful details of decomposition of arbitrary functions on the harmonic sphere $SU(2)/U(1)$ into infinite series of irreducible isospin components and establish the precise
correspondence between the relevant projection operators and their more accustomed representation via infinite products.

\section{Definitions and notations}


\subsection{Basic conventions}

The space-time tensors are marked either by the four-vector indices $m = 0,..,3$ or by a pair of spinor indices  $\alpha\dot{\alpha}; \,\alpha = 1,2; \dot{\alpha} = 1,2$. The two notations
are related via
\begin{equation}
x^{\alpha\dot{\alpha}} = (\tilde\sigma_m)^{\alpha\dot{\alpha}}x^{m}\,,  \;\; \partial_{\alpha\dot{\alpha}} = \frac{1}{2}(\sigma^m)_{\alpha\dot{\alpha}}\partial_{m}\,, \quad
\partial_{\alpha\dot{\alpha}} := \frac{\partial}{\partial x^{\alpha\dot{\alpha}}}, \quad \partial_{m} :=\frac{\partial}{\partial x^m}\,,
     \label{Def1}
\end{equation}
where
\begin{equation}\label{def/sigmasymbol}
(\sigma^m)_{\alpha\dot{\alpha}} = (\delta_{\alpha\dot{\alpha}},
\sigma^{i}_{\alpha\dot{\alpha}}), \qquad (\tilde{\sigma}^m)^{\dot{\alpha}\alpha} = (\delta^{\dot{\alpha}\alpha}, -(\sigma^{i})^{\dot{\alpha}\alpha})
= \varepsilon^{\alpha\beta}\varepsilon^{\dot\alpha\dot\beta}(\sigma^m)_{\beta\dot{\beta}}\,,
\end{equation}
and $\sigma^{i}_{\alpha\dot{\alpha}}$ are Pauli matrices. Spinor indices are raised  and lowered by the constant antisymmetric bi-spinors $\varepsilon_{\alpha\beta}, \varepsilon^{\alpha\beta}$,
such that $\varepsilon_{12} = -\varepsilon_{21} = 1$ and
\begin{equation}
    \varepsilon^{\alpha\beta}\varepsilon_{\beta\gamma} = \delta^{\alpha}_{\gamma} \nonumber
\end{equation}
(and the same for dotted indices). Sometimes we will use the condensed notation:
\begin{equation}
   ( \theta^i_{\alpha}, \bar{\theta}^i_{\dot{\alpha}}) \quad \leftrightarrow \quad \theta^i_{\hat{\alpha}}\,, \quad \hat{\alpha} = (\alpha, \dot\alpha).
\end{equation}
Our other  conventions are $\eta_{mn} = {\rm diag} (1, -1, -1, -1)$ and $\varepsilon_{0123} = 1$. We will use the definition
\be
\Box = \partial^n\partial_n = 2 \partial^{\alpha\dot\alpha}\partial_{\alpha\dot\alpha}\,. \lb{Box}
\ee

\subsection{Supersymmetry}
The generators of the even space-time part of the ${\cal N}$ extended $4D$ Poincar\'e supersymmetry algebra, viz. $4D$ Poincar\'e algebra (in the spinor notation),
satisfy the commutation relations:
\begin{eqnarray}
&&        [M_{\alpha\beta}, P_{\gamma\dot{\gamma}}] = i\varepsilon_{\gamma(\alpha}P_{\beta)\dot{\gamma}}\,,
        \qquad [\bar{M}_{\dot{\alpha}\dot{\beta}}, P_{\gamma\dot{\gamma}}] = i\varepsilon_{\dot{\gamma}(\dot{\alpha}}P_{\dot{\beta})\gamma}\,, \nonumber\\
&&        [M_{\alpha\beta}, M_{\gamma\xi}] = i\varepsilon_{\gamma(\alpha}M_{\beta)\xi} + i\varepsilon_{\xi(\alpha}M_{\beta)\gamma}\,, \nonumber\\
&&        [\bar{M}_{\dot{\alpha}\dot{\beta}}, \bar{M}_{\dot{\gamma}\dot{\xi}}] = i\varepsilon_{\dot{\gamma}(\dot{\alpha}}\bar{M}_{\dot{\beta})\dot{\xi}} + i\varepsilon_{\dot{\xi}(\dot{\alpha}}\bar{M}_{\dot{\beta})\dot{\gamma}}\,, \nonumber\\
&&        [M_{\alpha\beta}, \bar{M}_{\dot{\alpha}\dot{\beta}}] = 0\,,\quad [P_{\alpha\dot{\alpha}}, P_{\gamma\dot{\gamma}}] = 0\,,\label{def/even_algebra}
\end{eqnarray}
while the relations with the odd generators $Q^i_\alpha, \bar{Q}_{\dot\alpha j}$ and the $U(\mathcal{N})$ automorphism generators $A^i_j$ are given by
\begin{eqnarray}
&&        \{Q^i_{\alpha}, \bar{Q}_{\dot{\alpha}j}\} = 4 \delta^i_j P_{\alpha\dot{\alpha}}\,,  \nonumber \\
&&        [M_{\alpha\beta}, Q^i_{\gamma}] = i\varepsilon_{\gamma(\alpha}Q^i_{\beta)}\,, \qquad [\bar{M}_{\dot{\alpha}\dot{\beta}}, Q^i_{\alpha}] = 0\,, \nonumber \\
&&        [\bar{M}_{\dot{\alpha}\dot{\beta}}, \bar{Q}^i_{\dot{\gamma}}] = i\varepsilon_{\dot{\gamma}(\dot{\alpha}}\bar{Q}^i_{\dot{\beta})}\,,
\qquad [M_{\alpha\beta}, \bar{Q}^i_{\dot{\alpha}}] = 0\,, \label{def/odd_algebra}
\end{eqnarray}
and
\begin{eqnarray}
&&        [A^i_j , Q^k_{\alpha}] = -i\delta^k_j Q^i_\alpha\,, \qquad [A^i_j, \bar{Q}_{\dot{\alpha}k}] = i\delta^i_k\bar{Q}_{\dot{\alpha}j}\,, \nonumber\\
&&        [A^i_j,A^k_l] = i(\delta^i_lA^k_j- \delta_j^kA^i_l)\,.\label{def/odd_algebra2}
\end{eqnarray}
After splitting $U(\mathcal{N})$ into $SU(\mathcal{N}) \times U(1)$,
\begin{equation}\label{def/R+T_su2}
     A^i_i := R, \qquad  A^i_j := T^i_j + \frac{1}{\mathcal{N}}\delta^i_j R,
\end{equation}
the relations \eqref{def/odd_algebra2} are rewritten as:
\begin{eqnarray}
&&   [T^i_j,T^k_l] = i(\delta^i_lT^k_j- \delta_j^kT^i_l), \quad [R, T^i_j] = 0\,, \nonumber\\
&&    [T^i_j , Q^k_{\alpha}] = -i\left(\delta^k_j \delta^i_l - \frac{1}{\mathcal{N}} \delta^i_j \delta^k_l  \right)Q^l_\alpha\,, \quad  [T^i_j, \bar{Q}_{\dot{\alpha}k}]
= i\left(\delta^i_k \delta^l_j - \frac{1}{\mathcal{N}} \delta^i_j \delta^l_k \right)\bar{Q}_{\dot{\alpha}l}, \nonumber\\
&&    [R , Q^k_{\alpha}] = -iQ^k_\alpha, \quad [R, \bar{Q}_{\dot{\alpha}k}] = i\bar{Q}_{\dot{\alpha}k}\,. \label{RTrel}
\end{eqnarray}

\subsection{Superspace and superfields}
The realization of supertranslations in the standard superspace,
\begin{equation}\label{def/superspace}
    \mathbb{R}^{4|4\mathcal{N}} = (x^{\alpha{\dot{\alpha}}}, \theta^{\alpha}_i, \bar{\theta}^{\dot{\alpha} j}), \qquad i,j = 1,...,\mathcal{N},
\end{equation}
where $\theta^{\alpha}_i, \bar{\theta}^{\dot{\alpha}}$ are Grassmann variables, is given by:
\begin{equation}\label{def/supertranslation}
\delta x_{\alpha\dot{\alpha}} = 2i \big(\epsilon^i_\alpha \bar{\theta}_{\dot{\alpha}i} - \theta^i_\alpha \bar{\epsilon}_{\dot{\alpha}i}\big),  \qquad \delta \theta^\alpha_i
= \epsilon_i^\alpha \qquad \delta \bar{\theta}_{\dot{\alpha} i} = \bar{\epsilon}_{\dot{\alpha}i}\,.
\end{equation}

The superfields $\Phi_{\cal A}$ are given on the superspace (\ref{def/superspace}) and can carry some multi-index ${\cal A}$ of the Lorentz and $U(N)$ groups.
The supertranslations (\ref{def/supertranslation})
act on  them  according to the rule:
\begin{equation}
    \delta_{\epsilon} \Phi_{\cal A} \simeq \Phi'_{\cal A}(x', \theta', \bar{\theta}') - \Phi_{\cal A}(x, \theta, \bar{\theta}) = 0\,.
\end{equation}
Other transformations of the Poincar\'e supersymmetry in superspace correspond to the proper realizations of the generators (\ref{def/even_algebra}), (\ref{def/odd_algebra})
\footnote{Generically, the generators ${S_t}_{\cal A}^{\cal B}$ are red off from the
 ``active'' form of the transformations, $\Phi'_{\cal A}(Z) \simeq \Phi_{\cal A}(Z) - \delta Z \frac{\partial}{\partial_Z}\Phi_{\cal A}(Z) + M_{\cal A}^{\cal B}\Phi_{\cal B}(Z) :=
 \Phi_{\cal A}(Z) + i\omega_t{S_t}_{\cal A}^{\cal B}\,\Phi_{\cal B}(Z)$, where  $M_{\cal A}^{\cal B}$ is a matrix part of the (super)symmetry transformation and $\omega_t$
 is the relevant infinitesimal parameter.}. All they are divided into two parts, viz.
 the \textit{matrix} part acting on the external indices of $\Phi_{\cal A}$, and the \textit{coordinate} differential part acting on the superspace coordinates.
 In  its turn, the latter is split into the two parts as well, the one acting on Minkowski coordinates $x^m$ and another acting on $\theta$.

For example, the Lorentz generators  $M_{\alpha\beta}$,   $\bar{M}_{\dot\alpha\dot\beta}$ split in the following way:
\begin{eqnarray}\label{def/Jpq}
M_{\alpha\beta} =  {\cal M}_{\alpha\beta} + M^{x}_{\alpha\beta} + M^{\theta}_{\alpha\beta}\,, \qquad
\bar{M}_{\dot\alpha\dot\beta} =  \bar{{\cal M}}_{\dot\alpha\dot\beta} + M^{x}_{\alpha\beta} + \bar{M}^{\theta}_{\dot\alpha\dot\beta}\,,
\end{eqnarray}
where
\begin{eqnarray}
&& M^{x}_{\alpha\beta} = -ix_{\dot{\gamma}(\alpha} \partial^{\dot{\gamma}}_{\beta)}\,, \quad M^{\theta}_{\alpha\beta} = i\theta_{i(\alpha} \partial^{i}_{\beta)}\,, \quad
\partial^{i}_{\beta} := \frac{\partial}{\partial \theta^\beta_i}\,, \lb{Hol} \\
&& \bar{M}^{x}_{\dot\alpha\dot\beta} = ix_{(\dot{\alpha}}^\gamma \partial_{{\gamma}\dot{\beta})}\,, \quad \bar{M}^{\theta}_{\dot\alpha\dot\beta} = i\bar\theta^{i}_{(\dot\alpha}
\bar\partial_{i \dot{\beta})}\,, \quad \bar\partial_{i \dot\beta} := \frac{\partial}{\partial \bar{\theta}^{i \dot{\beta}}}\,.\lb{AHol}
\end{eqnarray}
The action of the matrix (spin) part ${\cal M}_{\alpha{\beta}}$ on the external doublet index $\gamma$ of the superfield $\Phi_\gamma$ is  defined as
    \begin{equation}
        ({\cal M}_{\alpha\beta})_\gamma^\nu \Phi_{\nu} = -i\varepsilon_{\gamma(\alpha} \Phi_{\beta)}\, \; \rightarrow  \;({\cal M}_{\alpha\beta})_\gamma^\nu = -i\varepsilon_{\gamma(\alpha}\delta^\nu_{\beta)}\,.\lb{LorMatr}
    \end{equation}
The action of $\bar{{\cal M}}_{\dot\alpha\dot\beta}$ on $\tilde{\Phi}_{\dot\gamma}$ is obtained via the complex conjugation,
\be
(\bar{{\cal M}}_{\dot\alpha\dot\beta})_{\dot\gamma}^{\dot\nu} = -i\varepsilon_{\dot\gamma(\dot\alpha}\delta^{\dot\nu}_{\dot\beta)}\,. \lb{LorMatr2}
\ee
Note that the signs in \eqref{LorMatr}, \eqref{LorMatr2} are chosen so that the Lorentz rotations with respect to the external doublet indices of the superfields
agree with those of $\theta^\alpha_i, \bar\theta^{\dot\alpha i}$. The spin Lorentz matrices $({\cal M}_{\alpha{\beta}})^\nu_\gamma$
and $(\bar{{\cal M}}_{\dot\alpha\dot\beta})_{\dot\gamma}^{\dot\nu}$
defined in  \eqref{LorMatr}, \eqref{LorMatr2} can be checked to obey the same commutation relations as $M_{\alpha{\beta}}$ and $M_{\dot\alpha\dot\beta}$ in \eqref{def/even_algebra}.
The generalization  of these rules to
a general superfield which is irreducible with respect to the external Lorentz index, i.e. $\Phi_{(\gamma_1\cdots \gamma_n)(\dot\nu_1\cdots \dot\nu_m)}$,
is straightforward\footnote{The action of the Lorentz generators on some field $\Phi_L$,
where $L$ stands for some Lorentz multi-index, is defined as $(M_{\alpha\beta})_L^N \Phi_N, \quad ({M}_{\alpha\beta})_L^N  = ({\cal M}_{\alpha\beta})_L^N + \delta^N_L \big(M^{x}_{\alpha\beta} + M^{\theta}_{\alpha\beta}\big)$
(and c.c.).}.

The $SU(\mathcal{N})$ and $U(1)$ generators are split in a similar way:
\begin{eqnarray}
&&    T^i_j = {\cal T}^i_j + (T^\theta)^i_j, \lb{SplInt} \\
&&    (T^\theta)^i_j = i\theta^\alpha_j \partial^i_\alpha + i\bar{\theta}^i_{\dot{\alpha}}\bar{\partial}^{\dot{\alpha}}_j
- \frac{i}{\mathcal{N}}\delta^i_j \left(\theta^\alpha_k \partial^k_\alpha + \bar{\theta}^k_{\dot{\alpha}}\bar{\partial}^{\dot{\alpha}}_k\right), \lb{thetaT}\\
&&   ({\cal T}^i_j)_k^l \Phi_{l} = -i\left(\delta^i_{k}\delta^{l}_{j} - \frac{1}{\mathcal{N}}\delta^{i}_j \delta^{l}_{k}\right)\Phi_{l}\,,  \label{def/Tgens} \\
&&    R = {\cal R} + R^\theta, \lb{SplR}\\
&&    R^\theta = i\theta^\alpha_k \partial^k_\alpha + i\bar{\theta}^k_{\dot{\alpha}}\bar{\partial}^{\dot{\alpha}}_k\,, \lb{thetaR}\\
&&   {\cal R}\Phi = i\lambda_R\Phi, \label{Rgen}
\end{eqnarray}
$\lambda_R$ being the $R$-charge of a complex superfield $\Phi$.
The rest of the supergroup generators are expressed as:
\begin{eqnarray}
&&    P_{\alpha\dot{\alpha}} = -i\partial_{\alpha\dot{\alpha}}, \quad Q^i_{\alpha} = i\partial^i_{\alpha} + 2\bar{\theta}^{i\dot{\alpha}}\partial_{\alpha\dot{\alpha}}, \quad
    \bar{Q}_{i\dot{\alpha}}, = -i\bar{\partial}_{i\dot{\alpha}}-2\theta^{\alpha}_i\partial_{\alpha\dot{\alpha}},\label{def/PQbarQ} \\
&& \{Q^i_{\alpha},\bar{Q}_{j\dot{\alpha}}\} = -4i\delta^i_j \partial_{\alpha\dot\alpha}\,, \quad \{Q^i_{\alpha}, {Q}^{j}_{\beta}\} = 0\,.\lb{AnticComm1}
\end{eqnarray}

The covariant spinor derivatives are defined by the standard expressions,
\begin{eqnarray}
&&    D^k_\alpha = \partial^k_\alpha + 2i\bar{\theta}^{k\dot{\alpha}}\partial_{\alpha\dot{\alpha}}, \quad
    \bar{D}_{k \dot{\alpha}} = -\bar{\partial}_{k\dot{\alpha}} - 2i\theta_{k}^{\alpha}\partial_{\alpha\dot{\alpha}},\label{def/spincovdercentral} \\
&&    \{D^k_{\alpha}, \bar{D}_{j\dot{\alpha}}\} = -4i \delta^k_j\partial_{\alpha\dot{\alpha}}\,, \quad \{D^k_{\alpha}, {D}^{j}_{\beta}\} = 0\,.  \label{def/spincovdercommutator}
\end{eqnarray}
They anticommute with the supertranslation generators:
\begin{equation}
 \{D^i_{\alpha}, \bar{Q}_{j\dot{\alpha}}\} =  \{D^i_{\alpha}, {Q}^{j}_{\beta}\} = 0\,.
\end{equation}
As a result, $D^i_{\alpha} \Phi$ and $\bar{D}^{\dot{\alpha}}_i \Phi$ are superfields having the correct transformation properties with respect to all generators of ${\cal N}$-extended
Poincar\'e supergroup.

\subsection{Harmonic superspace}

Hereafter we focus our attention on $\mathcal{N}=2$ harmonic superspace (HSS) \cite{Book}. In this superspace, superfields are functions of standard 4D superspace coordinates
$(x, \theta, \bar{\theta})$, as well as of additional $SU(2)/U(1)$ \textit{harmonic variables} $u^{\pm}_i, i=1,2$, satisfying the unitarity constraint
\begin{equation}
u^{+i}u^{-}_i = 1,
\end{equation}
or, equivalently, the completeness relation
\begin{equation}
    u^{+i}u^{-}_j - u^{-i}u^{+}_j = \delta^i_j\,.
\end{equation}

One of the parametrizations of HSS is the {\it central basis},
\begin{equation}
Z^{(C)} := (x_{\alpha\dot{\alpha}}, \theta^{\alpha}_i, {\bar{\theta}}^{\dot{\alpha} i},  u^{\pm}_i),
\end{equation}
which is just an extension of (\ref{def/superspace}) by harmonic variables. In addition to the standard spinor covariant derivatives
(\ref{def/spincovdercentral}), there are now harmonic covariant derivatives:
\begin{eqnarray}
&&D^{\pm\pm} = \partial^{\pm\pm}, \quad D^{0} = \partial^{0}, \\
&& \partial^{\pm\pm} = u^{\pm i} \frac{\partial}{\partial u^{\mp i}}, \qquad \partial^0 = u^{+i}\frac{\partial}{\partial u^{+i}} - u^{-i}\frac{\partial}{\partial u^{-i}}.
\end{eqnarray}
They satisfy the $su(2)$ algebra relations:
\begin{equation}
[{D^{++}},{D^{--}}] = D^0, \qquad [D^0, {D^{\pm\pm}}] = \pm 2 D^{\pm\pm}
\end{equation}
and commute with all the differential operators acting on ${\cal N}=2$ superspace and defined in the previous subsections.

The main peculiarity of HSS is the existence of another set of coordinates forming the {\it analytic basis}:
\begin{equation}
    Z^{(A)} := (x^{(A)}_{\alpha\dot{\alpha}}, {\theta^{(A)}}^{+}_{\hat{\alpha}}, {\theta^{(A)}}^{-}_{\hat{\alpha}}, u^{\pm}_i)\,, \qquad \hat{\alpha} :=(\alpha, \dot\alpha)\,.
\end{equation}
The two bases are connected by the following change of coordinates:
\begin{equation}\label{Baschange}
x^{(A)}_{\alpha\dot{\alpha}} = x_{\alpha\dot{\alpha}} - 4i\theta^{(i}_\alpha\theta^{j)}_{\dot{\alpha}} u^+_i u^-_j, \quad
\theta^{\pm (A)}_{\hat{\alpha}} = u^{\pm}_i \theta^i_{\hat{\alpha}}\,.
\end{equation}
The analytic basis makes manifest the presence of an invariant analytic subspace in HSS, involving half the number of Grassmann variables,
\bea
\zeta := (x^{(A)}_{\alpha\dot{\alpha}}, {\theta^{(A)}}^{+}_{\hat{\alpha}},u^{\pm}_i)\,.\lb{AnalSS}
\eea
The ${\cal N}=2$ supersymmetry acts on this set of  coordinates as
\bea
\delta_{\epsilon}x^{(A)\alpha\dot{\alpha}} = -4i(\epsilon^{i\alpha}\bar\theta^{+\dot\alpha} + \theta^{+\alpha}\bar\epsilon^{\dot\alpha i})u^-_i\,, \quad
\delta_{\epsilon}{\theta^{(A)}}^{+}_{\hat{\alpha}} = u^+_i\epsilon_{\hat{\alpha}}^i\,,  \quad \delta_{\epsilon}u^{\pm}_i = 0\,.
\eea
In what follows, we will mainly deal with the analytic basis and so drop the superscript $(A)$ on various quantities.

The spinor covariant derivatives in the central basis (\ref{def/spincovdercentral}) can be replaced by  their harmonic projections:
\begin{equation}\label{def/spincovderanalytic}
    D^{\pm}_\alpha = u^{\pm}_i D^{i}_\alpha\,, \qquad \bar{D}^{\pm}_{\dot{\alpha}} = u^{\pm}_i \bar{D}^{i}_{\dot{\alpha}}\,.
\end{equation}
After passing to the analytic basis these projections acquire the  following explicit form
\begin{subequations}
    \begin{align}
    D^+_{\hat{\alpha}} &= \partial^+_{\hat{\alpha}}, \quad \partial^+_{\hat{\alpha}} := \frac{\partial}{\partial \theta^{-\hat\alpha}}\,,  \lb{Short} \\
    D^-_{\alpha} &= -\partial^-_\alpha + 4i\bar{\theta}^{-\dot{\alpha}}\partial_{\alpha\dot{\alpha}}, \quad
    \bar{D}^-_{\dot{\alpha}} = -\partial^-_{\dot{\alpha}} - 4i{\theta}^{-\alpha}\partial_{\alpha\dot{\alpha}}, \quad \partial^-_{\hat{\alpha}} :=
    \frac{\partial}{\partial \theta^{+\hat\alpha}}\,.
    \end{align}
\end{subequations}

Using (\ref{def/spincovderanalytic}) and (\ref{def/spincovdercommutator}) one can derive the relations:
\begin{subequations}\label{def/spincovdercommutatoranalyt}
    \begin{align}
    \{D^{\pm}_{\hat{\alpha}},D^{\pm}_{\hat{\beta}}\} &= \{D^+_{\alpha}, D^-_\beta\} = \{\bar{D}^+_{\dot{\alpha}},\bar{D}^-_{\dot{\beta}}\} = 0, \\
    \{D^{+}_{\alpha},\bar{D}^{-}_{\dot{\alpha}}\} &= -\{\bar{D}^{+}_{\dot{\alpha}},{D}^{-}_{\alpha}\} = -4i\partial_{\alpha{\dot{\alpha}}}\,.
    \end{align}
\end{subequations}

The covariant harmonic derivatives in the analytic basis are obtained from those in the central basis also through the variable change \eqref{Baschange}:
\begin{eqnarray}
&&    D^{++} = \partial^{++} -4i\theta^{+\alpha}\bar{\theta}^{+\dot{\alpha}}\partial_{\alpha\dot{\alpha}} + \theta^{+\hat{\alpha}}\partial^+_{\hat{\alpha}}, \quad
D^{--} = \partial^{--} -4i\theta^{-\alpha}\bar{\theta}^{-\dot{\alpha}}\partial_{\alpha\dot{\alpha}} + \theta^{-\hat{\alpha}}\partial^-_{\hat{\alpha}}, \lb{D++--} \\
&& D^0 = \partial^{0} + \theta^{+\hat{\alpha}}\partial^-_{\hat{\alpha}} - \theta^{-\hat{\alpha}}\partial^{+}_{\hat{\alpha}}. \lb{D0}
\end{eqnarray}
They have the following commutation relations with the spinor derivatives:
\begin{eqnarray}
[D^{\pm\pm},D^{\pm}_{\hat{\alpha}}] = 0\,, \quad
        [D^{\pm\pm},D^{\mp}_{\hat{\alpha}}] = D^{\pm}_{\hat{\alpha}}\,, \quad
        [D^{0},D^{\pm}_{\hat{\alpha}}] = \pm D^{\pm}_{\hat{\alpha}} \,.\lb{HarSpin}
\end{eqnarray}

The functions $f^{+q}$ defined on the harmonic sphere $S^2 \sim SU(2)/U(1)$ are assumed to have a \textit{definite} harmonic
$U(1)$ charge $q$,
\be
D^0 f^{+q} = q\,f^{+q}\,.
\ee
An expansion of $f^{+q}$ over harmonic monomials should respect this condition\footnote{Actually, on $S^2$ one can define two equivalent sets of the harmonic functions (or ``sections'' ),
with integer $U(1)$ charges $|q|$ and
    $-|q|$, both giving rise to the same isospin contents. Hereafter, we consider the sector with non-negative  $U(1)$ charges (the notation $+q$ means $q$ harmonic charges $+$),
    since in the supersymmetric case just this option is compatible with ${\cal N}=2$ Grassmann harmonic analyticity defined
    in the standard way \cite{Book}.}. For instance, for the choice $q=1$ we have the following harmonic expansion
\begin{equation}
    f^{+}(u) = f^{i}u^+_i + f^{(ijk)}u^+_i u^+_j u^-_k + ...\,.
\end{equation}
In accord with this rule, the harmonic superfields $\Phi^{+q}(Z)$ in their $\theta$ expansion involve components which are the  harmonic functions with the appropriate fixed charges, e.g., for $q=1$:
\begin{equation}\label{superfield_expansion}
    \Phi^+(x, \theta^{+}, \theta^-, u) = \phi^{+}(x, u) + \theta^{+\hat{\alpha}} \psi_{\hat{\alpha}}(x, u) + \theta^{-\hat{\alpha}} \kappa^{++}_{\hat{\alpha}}(x, u) + ...\,.
\end{equation}
For general $q\geq 0$:
\bea
D^0\Phi^{+q}(Z) = q\Phi^{+q}(Z) \Rightarrow \Phi^{+q}(Z) = \phi^{+q}(x, u){+}\theta^{+\hat{\alpha}} \psi_{\hat{\alpha}}^{+(q-1)} (x, u)
{+} \theta^{-\hat{\alpha}} \kappa^{+(q +1)}_{\hat{\alpha}}(x, u){+}\ldots. \lb{Generq}
\eea

Generically, the harmonic superfields can have any amount of external indices, both Lorentz ones and those of $SU(2)$. The presence of harmonics
allows one to convert any external $SU(2)$ index into an extra harmonic $U(1)$ charge. For instance, the isodoublet superfield $\Phi^{i+}$  amounts
to the two harmonic superfields, with the charges $q=2$ and $q=0$:
\begin{equation}\label{def/noisoindexes}
    \Phi^{i+}(x, \theta, u) \quad \rightarrow \quad \Phi^{++} = u^{+}_i \Phi^{i+}, \quad \Phi^{0} = u^{-}_i \Phi^{i+}.
\end{equation}
So we can confine our attention only to the harmonic superfields without external $SU(2)$ indices. Accordingly,
the matrix parts of the $SU(2)$ generators (for ${\cal N}=2$ in (\ref{def/Tgens})) in the realization on harmonic superfields
in the analytic basis are turned into the differential operators acting only on the harmonic variables:
\begin{equation}\label{T-on-u}
\begin{split}
{\cal T}^i_k \;\Rightarrow \; (T^{u})^i_k = iu_k^{+}\frac{\partial}{\partial u^{+}_i}
+iu_k^{-}\frac{\partial}{\partial u^{-}_i} - \frac{i}{2}\delta^{i}_k \left(u_l^{+}\frac{\partial}{\partial u^{+}_l} + u_l^{-}\frac{\partial}{\partial u^{-}_l}\right).
\end{split}
\end{equation}

The crucial feature of the analytic basis is the possibility to define the ``short'' harmonic superfields, the analytic superfields $\varphi^{+q}(\zeta)$,
which are defined on the analytic subspace \eqref{AnalSS} and so depend only on half of the original Grassmann coordinates, viz., $\theta^{+}, \bar{\theta}^{+}$.
 This property amounts to imposing the following covariant constraints:
\begin{equation}\label{analyticsuperfield}
    D^{+}_{{\alpha}}\Phi^{+q} =  {\bar D}^{+}_{\dot{\alpha}}\Phi^{+q} = 0\,, \quad D^0\Phi^{+q} = q \Phi^{+q}\,.
\end{equation}
Because of the shortness property \eqref{Short} of the involved  spinor derivatives, eqs. \eqref{analyticsuperfield} can be interpreted as the Grassmann
analyticity conditions 
intended just to kill the dependence on the second half of the Grassmann coordinates, i.e. on $\theta^{- \hat\alpha}$:
\be
\Phi^{+q}(Z) \rightarrow  \varphi^{+q}(\zeta)\,.
\ee

Since the harmonic derivative $D^{++}$ commutes with the spinor derivatives $D^{+}_{\hat{\alpha}}$, an important property is that the superfield $D^{++}\varphi^{+q}$ is analytic
like $\varphi^{+q}(\zeta)$ itself.

For further use, we also present here the explicit form of the ${\cal N}=2$ supersymmetry generators in the analytic basis:
\bea
&& Q^k_\alpha = -i\big(u^{+k}\partial^-_\alpha + u^{-k}\partial^+_\alpha \big) -4 u^{-k}\bar\theta^{+\dot{\beta}}\partial_{\alpha\dot\beta}, \;
\bar{Q}_{k\dot\alpha} = -i\big(u^+_{k}\partial^-_{\dot\alpha} + u^{-}_{k}\partial^+_{\dot\alpha} \big) +4 u^{-}_{k}\theta^{+\beta}\partial_{\beta\dot\alpha}, \lb{AnQ} \\
&& [D^{\pm\pm},  Q^i_{\hat\alpha}] =[D^{0},  Q^i_{\hat\alpha}] = 0\,.\lb{AnQD}
\eea

\subsection{Central charges}
The central-charge modified ${\cal N}=2,\, 4D$ superalgebra corresponds to the following modification of the supercharges $Q_\alpha^i, \bar Q_{\dot\alpha i}$ and the covariant derivatives:
\bea
&& Q_\alpha^i \;\rightarrow \; Q_\alpha^{(c) i} = Q_\alpha^i - \theta_\alpha^i \frac{\partial}{\partial x^5}\,, \quad \bar{Q}_{\dot\alpha j} \;\rightarrow \; \bar{Q}_{\dot\alpha j}^{(c)}
 = \bar{Q}_{\dot\alpha j}- \bar\theta_{\dot\alpha j}\frac{\partial}{\partial x^5}\,,\lb{CCmodQ} \\
&& D_\alpha^i \;\rightarrow \; D_\alpha^{(c) i} = D_\alpha^i - i\theta_\alpha^i \frac{\partial}{\partial x^5}\,, \quad
\bar{D}_{\dot\alpha j} \;\rightarrow \; \bar{D}_{\dot\alpha j}^{(c)}
 = \bar{D}_{\dot\alpha j}- i\bar\theta_{\dot\alpha j}\frac{\partial}{\partial x^5}\,,\lb{CCmodD} \\
&&\{D_\alpha^{(c) i}, D_\beta^{(c) k}\} = -2i\varepsilon_{\alpha\beta}\varepsilon^{ik}\,\frac{\partial}{\partial x^5}\,, \quad
\{\bar{D}_{\dot\alpha j}^{(c)}, \bar{D}_{\dot\beta k}^{(c)}\} = -2i \varepsilon_{\dot\alpha\dot\beta}\varepsilon_{ik}\, \frac{\partial}{\partial x^5}\,.\lb{DopAnt}
\eea
The central charge can be used to generate, through the well known Scherk-Schwarz mechanism, mass terms for some analytic superfields, e.g.,
for the hypermultiplet $q^+(\zeta)$. Note that in general ${\cal N}=2$ superalgebra admits two independent central charges. For simplicity, we
restrict our attention to the option with one such charge. The ${\cal N}=2$ superalgebra with one central charge lacks the $U(1)$ $R$-symmmetry automorphisms,
which can formally be restored after adding the second central charge.

The HSS formalism for the central-charge extended ${\cal N}=2$ supersymmetry is defined basically in the  same way as without central charges \cite{Book}.
The analytic subspace in this case  involves some shifted $x^{5(A)}$ coordinate and is closed under the properly modified supersymmetry transformations.

\section{Casimir operators}

Casimir operators for $\mathcal{N}$-extended $4D$ Poincar\' e superalgebra in the standard superspace $\mathbb{R}^{4|4\mathcal{N}}$ were constructed in \cite{Rittenberg:1981cp}.
The harmonic superspace involves in addition the harmonics $u^{\pm}_i$ as coordinates.
Since harmonics are affected only by $SU(2)$ generators, the expressions for the superspin and  the $U(1)$ R-symmetry Casimir operators are expected not to change.  At the same time, the explicit
expressions for the superisospin Casimir operator are properly modified.

\subsection{Superspin Casimir operator}

To deduce the Casimir operators for Poincar\'e superalgebra, one starts from their expressions  through the superalgebra generators, re-express them in terms of derivatives with respect to the coordinates,
and finally recollect all terms so as to gain covariant derivatives.

In Minkowski space, the spin Casimir operator is constructed from
the Pauli-Lubanski vector\footnote{In the vector notation
it is $W^{n} = -\frac{1}{2}\epsilon^{nmpq}P_{m}J_{pq},$ and $J_{pq}
= \frac{i}{2}(\sigma_{pq})_{\alpha\beta}M^{\beta\alpha} - \frac{i}{2}(\tilde{\sigma}_{pq})_{\dot{\alpha}\dot{\beta}}\bar{M}^{\dot{\beta}\dot{\alpha}}.$}:
\begin{eqnarray}\label{superspin}
    C_S = 2 W^{\dot{\alpha}\alpha}W_{\alpha\dot{\alpha}}\,, \quad
    W_{\alpha\dot{\alpha}} = \partial_{\alpha\dot{\gamma}}{\bar M}^{\dot{\gamma}}_{\dot\alpha} -\partial_{\gamma\dot{\alpha}}M^{\gamma}_{\alpha}\,.
\end{eqnarray}

In superspace,
\begin{equation}
    [W_{\alpha\dot{\alpha}},Q^i_\beta] = i\partial_{\beta\dot{\alpha}}Q^i_\alpha -\frac{i}{2}\partial_{\alpha\dot\alpha}Q^i_\beta, \qquad [W_{\alpha\dot{\alpha}}, \bar{Q}^i_{\dot{\beta}}]
    = i\partial_{\alpha\dot{\beta}}{\bar Q}^i_{\dot{\alpha}} -\frac{i}{2}\partial_{\alpha\dot{\alpha}}{\bar Q}^i_{\dot{\beta}},\lb{CommWQ}
\end{equation}
and one is led to modify $W_{\alpha\dot{\alpha}}$ to $\hat{W}_{\alpha\dot{\alpha}}$ which commutes with the supercharges $Q, \bar{Q}$
\begin{eqnarray}\label{modifiedPL}
&& \hat{W}_{\alpha\dot{\alpha}} = W_{\alpha\dot{\alpha}} + \frac{1}{8}[Q^i_{\alpha},\bar{Q}_{i\dot{\alpha}}]
- \frac{1}{4\Box}\partial_{\alpha\dot{\alpha}}\partial^{\dot{\gamma}\gamma}\left[Q^i_{\gamma},\bar{Q}_{i\dot{\gamma}}\right], \\
&& [\hat{W}_{\alpha\dot{\alpha}}, Q^i_{\gamma}] = [\hat{W}_{\alpha\dot{\alpha}}, \bar{Q}_{i\dot{\gamma}}] = 0,
\end{eqnarray}
where $\Box = \partial^{n}\partial_n = 2\partial^{\dot{\alpha}\alpha}\partial_{\alpha\dot{\alpha}} $. The definition  of the superspin operator as
\be
\hat{C}_s = 2\hat{W}^{\dot{\alpha}\alpha}\hat{W}_{\alpha\dot{\alpha}} \lb{NewSuCa}
\ee
differs from the standard definition
\be
\tilde{C}_s = Z^2\,, \;\;Z_{\beta\dot\beta\, \alpha\dot\alpha} = \partial_{\beta\dot\beta}\tilde{W}_{\alpha\dot{\alpha}}
- \partial_{\alpha\dot\alpha}\tilde{W}_{\beta\dot{\beta}} = \varepsilon_{\beta\alpha}\,\partial^\gamma_{(\dot\beta}
\tilde{W}_{\gamma\dot\alpha)} +\varepsilon_{\dot\beta\dot\alpha}\,\partial^{\dot\gamma}_{(\beta}\tilde{W}_{\alpha)\dot{\gamma}}\,,\lb{AltDef}
\ee
where
\be
\tilde{W}_{\alpha\dot{\alpha}} = W_{\alpha\dot{\alpha}} + \frac{1}{8}[Q^i_{\alpha},\bar{Q}_{i\dot{\alpha}}]\,, \quad
[\tilde{W}_{\alpha\dot{\alpha}}, Q^i_{\hat{\gamma}}] = -\frac{i}{2} \partial_{\alpha\dot\alpha} Q^i_{\hat{\gamma}}\,. \lb{AltDef2}
\ee
Making use of the property
$$
\partial_{\alpha\dot\alpha}W^{\alpha\dot\alpha} = 0\,,
$$
we can represent
\be
\hat{W}_{\alpha\dot{\alpha}} = \tilde{W}_{\alpha\dot{\alpha}} - 2\frac{1}{\Box}\partial_{\alpha\dot\alpha}\,\partial^{\beta\dot\beta}\tilde{W}_{\beta\dot\beta}
\ee
and establish the precise relation between the two definitions of the superspin Casimir operators:
\be
\tilde{C}_s = Z^2 = \Box \tilde{W}^2 - 2 \partial^{\beta\dot\beta}\tilde{W}_{\beta\dot\beta} \,\partial^{\alpha\dot\alpha}\tilde{W}_{\alpha\dot\alpha} = \Box \, \hat{W}^2 =  \frac12\,\Box \hat{C}_s \,.
\ee
Note that the alternative definition of the superspin Casimir as in eqs. \eqref{NewSuCa}, \eqref{modifiedPL} is also admissible in other ${\cal N}$ cases and was exploited, e.g., in \cite{ArvMezTown}.

Similarly to $M_{\alpha\beta}, \bar{M}_{\dot\alpha\dot\beta}$, the operator $W_{\alpha\dot{\alpha}}$ can be split into the matrix part and the coordinate parts:
\begin{align}
    W_{\alpha\dot{\alpha}} &= {\cal W}_{\alpha\dot{\alpha}} +  W^{\theta}_{\alpha\dot{\alpha}} + W^{x}_{\alpha\dot{\alpha}}\,, \lb{SplModW}\\
    W^{x/\theta}_{\alpha\dot{\alpha}} &= \partial_{\alpha\dot{\gamma}} (\bar M^{x/\theta})^{\dot{\gamma}}_{\dot\alpha}
    -\partial_{\gamma\dot{\alpha}}(M^{x/\theta})^{\gamma}_{\alpha}\,, \quad {\cal W}_{\alpha\dot{\alpha}} =
    \partial_{\alpha\dot{\gamma}} \bar{\cal M}^{\dot{\gamma}}_{\dot\alpha}
    - \partial_{\gamma\dot{\alpha}}{\cal M}^{\gamma}_{\alpha}\,.\lb{CommModW}
\end{align}

It is easy to see that $ W^x_{\alpha\dot{\alpha}} = 0 $.
After substituting (\ref{def/Jpq}), using identities for (\ref{def/sigmasymbol}) and plugging expressions for $Q$ and $D$ (\ref{def/PQbarQ}), (\ref{def/spincovdercentral}) into \eqref{modifiedPL}, \eqref{SplModW},
the remaining differential part is written as:
\begin{align}\label{Wtheta}
W^{\theta}_{\alpha\dot{\alpha}} =  -\frac{1}{8}[D^{i}_{\alpha}, \bar{D}_{i\dot{\alpha}}] -\frac{1}{8}[Q^i_{\alpha},\bar{Q}_{i\dot{\alpha}}] + \frac{1}{4\Box}\left([D^{i}_{\gamma}, \bar{D}_{i\dot{\gamma}}]
+[Q^i_{\gamma},\bar{Q}_{i\dot{\gamma}}]\right)\partial^{\dot{\gamma}\gamma}\partial_{\alpha\dot{\alpha}}.
\end{align}
After plugging (\ref{Wtheta}) in (\ref{modifiedPL}) we get the final expression for $\hat{W}$:
\begin{align}
\hat{W}_{\alpha\dot{\alpha}} = {\cal W}_{\alpha\dot{\alpha}} -\frac{1}{8}[D^{i}_{\alpha}, \bar{D}_{i\dot{\alpha}}]
+ \frac{1}{4\Box}[D^{i}_{\gamma}, \bar{D}_{i\dot{\gamma}}]\partial^{\dot{\gamma}\gamma}\partial_{\alpha\dot{\alpha}}\,.
\end{align}
Being translated to the analytic basis with making use of the definitions (\ref{def/spincovderanalytic}) and the relation
\bea
[D^{i}_{\alpha}, \bar{D}_{i\dot{\alpha}}] = -2 \big(D^-_\alpha \bar{D}^+_{\dot\alpha} + \bar{D}^-_{\dot\alpha} D^+_\alpha \big), \lb{CommDD}
\eea
it takes the form
\begin{align}
\hat{W}_{\alpha\dot{\alpha}} = {\cal W}_{\alpha\dot{\alpha}} +\frac{1}{4}(D^-_{\alpha}\bar{D}^{+}_{\dot{\alpha}} + \bar{D}^-_{\dot{\alpha}}D^+_\alpha)
- \frac{1}{2\Box}(D^-_{\gamma}\bar{D}^{+}_{\dot{\gamma}} + \bar{D}^-_{\dot{\gamma}}D^+_\gamma)\partial^{\dot{\gamma}\gamma}\partial_{\alpha\dot{\alpha}}\,. \label{pauli_lubanski_final}
\end{align}
We observe that in application to the analytic superfields $\hat{W}_{\alpha\dot{\alpha}}$ is reduced to its matrix part ${\cal W}_{\alpha\dot{\alpha}}$. Therefore,
on any analytic superfield without Lorentz indices $\hat{W}_{\alpha\dot{\alpha}} =0\,, \;\hat{C}_s = \tilde{C}_s = 0$.

Notice that $\hat{W}_{\alpha\dot{\alpha}}$ does not commute with spinor derivatives:
\begin{equation}
[\hat{W}_{\alpha\dot{\alpha}}, D^{\pm}_{\beta}] = i\partial_{\beta\dot{\alpha}}D^{\pm}_{\alpha} - \frac{i}{2}\partial_{\alpha\dot{\alpha}}D^{\pm}_{\beta}\,, \quad
[\hat{W}_{\alpha\dot{\alpha}}, \bar{D}^{\pm}_{\dot\beta}] = -i\partial_{\alpha\dot{\beta}}\bar{D}^{\pm}_{\dot\alpha} + \frac{i}{2}\partial_{\alpha\dot{\alpha}}\bar{D}^{\pm}_{\dot\beta}\,.
\end{equation}
Given an analytic off-shell superfield $\Phi^{+q}$ and employing these relations, it is easy to determine spins of irreducible covariant superfield projections of $\Phi^{+q}$, namely, $\Phi^{+q},
D^-_{\hat\alpha}\Phi^{+q},D^-_{\alpha}\bar{D}^-_{\dot\alpha}\Phi^{+q}$, $(D^-)^2\Phi^{+q}$,
$(\bar{D}^-)^2\Phi^{+q}, (D^-)^4\Phi^{+q}, \bar{D}^-_{\dot\alpha}(D^-)^2\Phi^{+q}, D^-_{\alpha}(\bar{D}^-)^2\Phi^{+q}$
\footnote{The vector projection $D^-_{\alpha}\bar{D}^-_{\dot\alpha}\Phi^{q}$ should be
as usual split into irreducible 4-transverse and gradient pieces.}. Using $\hat{W}_{\alpha\dot{\alpha}}\Phi^{q} = 0$, it is straightforward to check that
$$
\hat{C}_s  \Phi^{q} = 0\,, \quad  \hat{C}_s (D^-_{\hat\alpha}\Phi^{q}) = \frac12\big(\frac12 +1\big)\,\Box  (D^-_{\hat\alpha}\Phi^{q})\,, \ldots \,.
$$

\subsection{Superisospin Casimir operator}

To construct the superisospin Casimir operator in the present setting, we start just with the algebra of $SU(2)_A$,
\begin{equation}
    [T^i_j, T^k_l] = i(\delta^i_l T^k_j - \delta^k_j T^i_l),
\end{equation}
and its Casimir:
\begin{equation}
    C_I = T^i_j T^j_i\,.\label{casimir_op}
\end{equation}
Here, the complete $SU(2)_A$ generators are
\begin{equation}
T^i_j = (T^\theta)^i_j + (T^{(u)})^i_j, \label{SumGen}
\end{equation}
where $(T^\theta)^i_j$ and $(T^{(u)})^i_j$ were defined in (\ref{def/Tgens}) and \eqref{T-on-u}.
Using the completeness relations for the harmonics, the second generator can be expressed through the  harmonic
derivatives,
\begin{equation}
T^{(u)ij} = iu^{+i}u^{+j} \partial^{--} - iu^{-i}u^{-j} \partial^{++}  - iu^{+(i}u^{j)-} \partial^0\,, \quad T^{(u)ij} = \varepsilon^{jk}(T^{(u)})^i_{k}\,.
\end{equation}
Then the pure harmonic part of the Casimir operator (\ref{casimir_op}) can be expressed in terms of the harmonic derivatives, so that its commutativity with $T^{(u)ij}$ becomes manifest:
\begin{equation}\label{casimir_op_su2}
    C_I = \frac{\partial_0^2}{4} + \frac{1}{2}\{\partial^{++}, \partial^{--}\}\,.
\end{equation}

Passing to harmonic superspace, one could guess that the superisospin operator is the same, but with the covariant harmonic derivatives
(independently of the choice of basis in superspace):
\begin{align}
(T^{(u)})^{ij} &= iu^{+i}u^{+j} D^{--} - iu^{-i}u^{-j} D^{++} - iu^{+(i}u^{j)-} D^0,\lb{TuExpan}
\end{align}
\begin{equation}\label{isospin_operator}
    \tilde{C}_I = \frac{(D^0)^2}{4} + \frac{1}{2}\{D^{++}, D^{--}\}\,.
\end{equation}
However, such a ``naive'' choice is not correct because the doublet $SU(2)_A$ indices are carried not only by the harmonic variables, but by Grassmann coordinates as well.
Respectively, the complete $SU(2)_A$ generators are given by the sum \eqref{SumGen} and one should define the $SU(2)_A$ Casimir through these complete generators.

The generators \eqref{SumGen} are still not appropriate for constructing such generic Casimir operator of ${\cal N}=2$ supersymmetry, since they do not commute with
the supercharges  $Q^i_{\hat\alpha}$,
\bea
[T^{(ik)}, Q_\alpha^j] = i\varepsilon^{j(i} Q^{k)}_\alpha\,, \quad [T^{(ik)}, \bar{Q}_{\dot\alpha}^{j}] =i\varepsilon^{j(i} \bar{Q}^{k)}_{\dot\alpha}\,. \lb{TQcomm}
\eea
It was observed in \cite{Rittenberg:1981cp,Gates} that one can construct the modified $SU(2)_A$ generators, such that they  {\it commute} with the supercharges:
\begin{subequations}
    \begin{align}
        &\hat{T}^i_j = T^i_j - \frac{1}{2\Box}[Q^{i\alpha},\bar{Q}^{\dot\alpha}_j]\partial_{\alpha\dot\alpha} +\frac{1}{4\Box}\delta^i_j[Q^{k\alpha},\bar{Q}^{\dot\alpha}_k]\partial_{\alpha\dot\alpha}, \label{Tij_modified} \\
        &[\hat{T}^i_j, Q^{k\alpha}] = [\hat{T}^i_j, \bar{Q}^{\dot\alpha}_k] = 0\,.
    \end{align}
\end{subequations}
Then the correct superisospin Casimir is defined as
\footnote{Though the ``would be'' Casimir operator \eqref{isospin_operator} also commutes with supercharges, it also commutes with both parts of the full $SU(2)$ generator \eqref{SumGen} separately
and so fully decouple. It yet can be used to specify the isospin contents of the massless on-shell analytic  supermultiplets (see sections 5, 6).}
\bea
C_I = \hat{T}^{i}_{j}\hat{T}^{j}_{i}.  \lb{SuperIso}
\eea

Now we can express $\hat{T}^i_j$ in terms of covariant derivatives. Using the expressions for $Q$ and $D$ (\ref{def/PQbarQ}), (\ref{def/spincovdercentral}), one can check
that
\begin{align}\label{Tprojection}
    \hat{T}^{(ik)}&= T^{(u)(ik)} +\frac{1}{2\Box}\left[D^{(i\alpha},\bar{D}^{j)\dot\alpha}\right]\partial_{\alpha\dot\alpha}\,.
\end{align}
Passing to the harmonic projections of the spinor covariant derivatives in \eqref{Tprojection}, and defining
\begin{equation}
    \hat{T}^{(ij)} = -iu^{-i}u^{-j} \nabla^{++} + iu^{+i}u^{+j}\nabla^{--}-iu^{+(i}u^{-j)}\nabla^{0},
\end{equation}
we find that the operators $\nabla^{\pm\pm}, \nabla^0$ are ``long'' versions of the harmonic covariant derivatives:
\begin{align}
    \nabla^{\pm\pm} = D^{\pm\pm} \pm \frac{i}{2\Box}[D^{\pm}_\alpha, \bar{D}^\pm_{\dot{\alpha}}]\partial^{\dot\alpha\alpha}, \quad \nabla^0
    = D^0 - \frac{i}{2\Box}([D^{-}_{\alpha},\bar{D}^{+}_{\dot\alpha}]  + [D^+_\alpha, \bar{D}^-_{\dot\alpha}])\partial^{\dot\alpha\alpha}\,.
\end{align}
With all $D^{+}_{\hat{\alpha}}$ spinor derivatives pulled out to the right, the modified harmonic derivatives take the form
\begin{eqnarray}
     &&   \nabla^{++} = D^{++} + \frac{i}{\Box}(D^{\alpha+}  \bar{D}^{+\dot\alpha})\partial_{\alpha\dot\alpha}, \nonumber \\
     &&   \nabla^{--} = D^{--} - \frac{i}{\Box}(D^{\alpha-}  \bar{D}^{-\dot\alpha})\partial_{\alpha\dot\alpha}, \nonumber \\
     &&   \nabla^{0} = (D^0 - 2) - \frac{i}{\Box}(D^{-}_{\alpha}\bar{D}^{+}_{\dot\alpha} - \bar{D}^-_{\dot\alpha}  D^+_\alpha)\partial^{\dot{\alpha}\alpha}\,. \label{Casimir/CovDerExpr/Nablas}
\end{eqnarray}
These new harmonic covariant derivatives allow one to bring the superisospin Casimir operator to the short form similar to (\ref{casimir_op_su2}):
\begin{equation}
C_I = \hat{T}^{i}_{j}\hat{T}^{j}_{i}  = \frac{1}{4}(\nabla^0)^2 + \frac{1}{2}\{\nabla^{++}, \nabla^{--}\}\,.
\end{equation}

The salient properties of the modified harmonic derivatives are, first, that they commute with {\it all} spinor covariant derivatives,
\begin{equation}
    [\nabla^{\pm\pm}, D^{+}_{\hat{\alpha}}] = [\nabla^{\pm\pm}, D^{-}_{\hat{\alpha}}] = [\nabla^{0}, D^{+}_{\hat{\alpha}}] = [\nabla^{0}, D^{-}_{\hat{\alpha}}] = 0
\end{equation}
and, secondly,  that they obey by themselves $SU(2)$ algebra, like their $D^{\pm\pm}, D^0$ prototypes,
\begin{equation}
[\nabla^{0}, \nabla^{\pm\pm}] = \pm 2\nabla^{\pm\pm}\,, \qquad [\nabla^{++}, \nabla^{--}] = \nabla^{0}\,. \lb{nablaAlg}
\end{equation}
They possess the following commutation relations with the ``old'' harmonic derivatives:
\begin{eqnarray}
  &&      [D^{\pm\pm},\nabla^{\mp\mp}] = \pm\nabla^0, \quad  [D^0, \nabla^{\pm\pm}] = \pm 2 \nabla^{\pm\pm}, \nonumber \\
  &&      [ D^{\pm\pm}, \nabla^0] = \mp 2\nabla^{\pm\pm}, \quad [D^0, \nabla^0] = 0\,. \lb{NablaD}
\end{eqnarray}

A curious peculiarity of the algebra \eqref{nablaAlg} is the universal shift by -2 of the new $U(1)$ charge operator  $\nabla^0$, which arose as a result  of
anticommuting spinor derivatives by the  relations (\ref{def/spincovdercommutatoranalyt}). This effective ``shift'' is present for any harmonic analytic superfield  and is responsible
for the correct superisospin contents of the analytic superfields \cite{Book}:

\begin{equation}\label{towerofisospins}
    \Phi^{(q)} (x_A, \theta^{+}, \bar{\theta}^+, u): \quad  I = \biggr|\frac{q-2}{2}\biggr| + n, \quad n = 0, 1, ... \phantom{1}
\end{equation}
(this formula holds both for $q\geq 0$ and for $q< 0$).

Finally, it is worth recalling that after passing to the analytic basis the full $SU(2)_A$ symmetry generator \eqref{SumGen} is reduced to its harmonic part,
\bea
[T^i_j]_{an} = (T^{(u)})^i_j\,, \lb{AnSU2}
\eea
which reflects the property that in this basis only the harmonic variables manifestly carry the $SU(2)_A$ (doublet) indices. All other relations
preserve their form, with the analytic-basis expressions \eqref{AnQ} for the ${\cal N}=2$ supersymmetry generators.

\subsection{U(1) charge Casimir operator}

This Casimir is constructed much similarly to the supeisospin Casimir; one can identify it with the modified  generator $\hat{R}$ commuting with the Q-supercharges,
\begin{align}
    &C_{U(1)} \sim \hat{R} = R - \frac{1}{2\Box}[Q^{i\alpha},\bar{Q}^{\dot\alpha}_i]\partial_{\alpha\dot\alpha}\,, \\
    &[\hat{R}, Q^i_{\hat{\alpha}}] = 0\,.
\end{align}
If we use the definitions of $Q$ and $D$ (\ref{def/PQbarQ}), (\ref{def/spincovdercentral}), as well as that of $R$ (eqs. (\ref{SplR}) and (\ref{thetaR})), we obtain
\begin{align}
\hat{R} &=  {\cal R}+ \frac{1}{2\Box}[D^{i}_{\alpha}, \bar{D}_{i\dot{\alpha}}]\partial^{\dot{\alpha}\alpha}\,.
\end{align}
In the analytic basis:
\begin{align}
C_{U(1)} := -i\hat{R} = -i{\cal R} -\frac{i}{\Box}(\bar{D}^-_{\dot\alpha}D^+_{\alpha} + D^{-}_{\alpha}\bar{D}^{+}_{\dot\alpha})\partial^{\dot\alpha\alpha}, \lb{U1Casimir}
\end{align}
where the factor $-i$ was introduced for further convenience.
Thus we observe that on the analytic superfields the Casimir $C_{U(1)}$  is reduced  to the pure ``matrix'' part ${\cal R}$ of $R$. This means that a complex analytic superfield can be characterized by some external $U(1)$
charge $\lambda_R$  associated with the $U(1)_R$ component of the full $U(2)_R = U(1)_R\times SU(2)_R$ $R$-symmetry (and different from the harmonic $U(1)$ charge).
All component fields of the given analytic superfield (including an infinite set of auxiliary fields) are eigenfunctions
of this ``external'' $U(1)$ Casimir with the same eigenvalue\footnote{To avoid a possible misunderstanding, we note that the component fields are generically transformed under $U(1)$ R-symmetry even for $\lambda_R=0$
because of the presence of the differential part in the $R$-charge generator. The coordinates $\theta^\pm_{\alpha}$ and $\bar\theta^\pm_{\dot\alpha}$ transform with opposite $U(1)_R$ phases and so do the associated
component fields in the $\theta$-expansions. So the component fields possess such ``intrinsic'' $U(1)$ R-symmetry charges in parallel with the external one  $\lambda_R$. }.

\section{Massive representations. Superisospin expansions}

As was noticed above, on the analytic superfields $\Phi^{+q}_{\cal A}$ (${\cal A}$ is an external  multi-index of the Lorentz group) two of the three Casimir operators are reduced to their matrix pieces,
\begin{equation}
C_S \Phi^{+q}_{\cal A} = 2 ({\cal W}^{\dot{\alpha}\alpha}{\cal W}_{\alpha\dot{\alpha}})_{\cal A}^{\cal B}\Phi^{+q}_{\cal B}, \qquad C_{U(1)} \Phi^{+q}_{\cal A }= -i{\cal R} \Phi^{+q}_{\cal A}\,,
\end{equation}
because terms with derivatives in these operators involve $D^{+}_{\hat{\alpha}}$ in the rightmost position and so vanish on the analytic superfields.

This implies that in order to split an analytic superfield into irreducible representations with fixed superspins and $U(1)_{R}$ charges one needs to perform
the same operations as those described, e.g., in \cite{Sokatchev:1975gg,Ponds} for the case of $\mathcal{N}=1$ superspace. Here we will not discuss the peculiarities of the superspin decomposition  and, instead,
focus on superfields without external spinor indices or  $U(1)_{R}$ charges. They all carry zero superspin.

So it is the superisospin operator $C_I$ which characterizes the irreducible representations of ${\cal N}=2$ supersymmetry in the case at hand.
Since there is an infinite tower of superisospins in the unconstrained analytic superfield (\ref{towerofisospins}), there should exist a projector for each superisospin,
as it is reflected in eq. (\ref{towerofisospins}):
\begin{equation}\label{decomposition_isospins}
    \begin{split}
        \Phi^{+q} &= {\Phi}^{+q}_{I=|\frac{q}{2} - 1|} + {\Phi}^{+q}_{I=|\frac{q}{2} - 1|+1} + {\Phi}^{+q}_{I=|\frac{q}{2} - 1|+2} + ...\\
        &= \text{П}^{(q)}_{I=|\frac{q}{2} - 1|}\Phi^{+q} + \text{П}^{(q)}_{I=|\frac{q}{2} - 1|+1}\Phi^{+q} + \text{П}^{(q)}_{I=|\frac{q}{2} - 1|+2}\Phi^{+q} + ...\,.
    \end{split}
\end{equation}

To find the expressions for the projection operators, one can apply the same techniques as in the Appendix A for the pure $SU(2)$ case, with replacing the derivatives
$\partial^{\pm}, \partial^{0}$ by $\nabla^{\pm\pm}, \nabla^0$. Note that the whole consideration below applies only to the superfields on which Box operator $\Box$ does not vanish
and so the inverse operator $\Box^{-1}$ is well defined.

The idea of decomposition over superisospins is as follows. Each superfield can be expressed in terms of its irreducible superisospin components (\ref{decomposition_isospins}).
These constituents are represented by the proper analytic superfields $\hat{\Phi}^{+q}$ constrained as,
\begin{equation}
    \nabla^{++}\hat{\Phi}^{+q} = 0, \qquad q \geq 2\,. \label{contraint_1}
\end{equation}
The value of superisospin can be properly determined for such irreducible superfields.
It should be pointed out that, because of the shift of $U(1)$ charge, $\nabla_0 = D_0 -2$,
there occurs an essential difference between the cases $q\geq 2$ and $q < 2$, i.e., $q=1$ and $q=0$. In the $q < 2$ case the condition \eqref{contraint_1} should be replaced by
\begin{equation}
    \nabla^{--}\hat{\Phi}^{+q} = 0,  \label{contraint_2}
\end{equation}
otherwise we would encounter non-physical ``negative superisospins''.

We postpone considering the special  $q=1, q=0$ cases to the end of this section and start with $q\geq 2$:
\begin{equation}
    C_I \hat{\Phi}^{+q} = \frac{q}{2}\cdot\frac{q-2}{2} \hat{\Phi}^{+q} \equiv I(I +1)\hat{\Phi}^{+q} \longrightarrow I = \frac{q-2}{2}\,.
\end{equation}
In order to relate an unconstrained analytic superfield $\Phi^{+q}$ to the constrained one (\ref{contraint_1}), we are led to construct the appropriate projection operator
from the analyticity-preserving harmonic derivatives $\nabla^{\pm\pm}$:
\begin{equation}
\hat{\Phi}^{+q}_{I=\frac{q-2}{2}} = \text{П} \Phi^{+q} = a_0 \Phi^{+q} + a_1 (\nabla^{--})(\nabla^{++}) \Phi^{+q} + a_2 (\nabla^{--})^2(\nabla^{++})^2 \Phi^{+q} + ...\,,
\end{equation}
where the numerical coefficients $a_k$ are undetermined at the moment. Since the projection operator must obey the property $\text{П}^2 = \text{П}$,
these  coefficients should be properly constrained, in particular, $a_0 = 1$. Since we want to satisfy the constraint (\ref{contraint_1}), all other coefficients also turn out to be properly fixed.
In the process, we apply the following general identity:

\begin{equation}
[\nabla^{++}, (\nabla^{--})^k] (\nabla^{++})^k \Phi^{+q} = k(q+k-1)(\nabla^{--})^{k-1}(\nabla^{++})^k \Phi^{+q}
\end{equation}
and then make use of the standard induction routine. In this way we obtain

\begin{equation}
    \hat{\Phi}^{+q}_{I=(q-2)/2} = \Phi^{+q} + \sum_{k=1}^{\infty} \frac{(-1)^k(q-1)!}{k! (k+q-1)!} (\nabla^{--})^{k}(\nabla^{++})^{k} \Phi^{+q}\,. \label{Irr-q}
\end{equation}
We observe that \eqref{Irr-q} differs from the pure $SU(2)$ expansion \eqref{A86} by shifting the number $m$ in the coefficients of the latter series by $-2$,
in accordance with the same shift of $\nabla^0$ compared to $D^0$.


Now, using the property that $\nabla^{\pm}$ does not change the value of superisospin, one can reverse the problem and express an unconstrained superfield ${\Phi}^{+q}$ as an infinite series over
all possible  constrained superfields with fixed superisospins of the type \eqref{Irr-q}:

\begin{eqnarray}\label{decomposition}
{\Phi}^{+q} = c_0 \hat{\Phi}^{+q}_{I=\frac{q-2}{2}} + c_1 (\nabla^{--})\hat{\Phi}^{+(q+2)}_{I=\frac{q}{2}}+ c_2 (\nabla^{--})^2\hat{\Phi}^{+(q+4)}_{I=\frac{q+2}{2}} + ... =
\sum_{k=0}^{\infty} c_k (\nabla^{--})^k\hat{\Phi}^{+(q+2k)}_{I=\frac{q}{2}+k-1}.
\end{eqnarray}

Here $\hat{\Phi}^{+(q+2k)}_{I=\frac{q}{2}+k-1}$ are expressed according to \eqref{Irr-q}, with the change
\bea
\Phi^{+q} \;\Rightarrow \; (\nabla^{++})^k \Phi^{+q}\,, \quad \nabla^{++}\hat{\Phi}^{+(q+2k)}_{I=\frac{q}{2}+k-1} =0 \,.
\eea

To determine the  coefficients $c_k$, one should substitute the expressions  \eqref{Irr-q} in \eqref{decomposition} and require all terms with derivatives
on the right-hand side of (\ref{decomposition}) to cancel among
themselves:
\begin{equation}\label{phi_expansion}
    \Phi^{+q} = (c_0) \Phi^{+q} + \left({c_1} - \frac{(q-1)!}{q!}c_0\right) (\nabla^{--}\nabla^{++}) \Phi^{+q} + ...\,.
\end{equation}
In this way  we obtain $c_0 = 1$, $c_1 = \frac{1}{q}$, etc. A direct computation of those coefficients is quite easy for the first
few terms and can be done by hand. To find the coefficients at any $k$ we could apply to the pure $SU(2)$ expansion \eqref{GenDec+m}.  Starting with $m\geq 2$, it evidently should coincide
with the expansion for $\Phi^{+q}$ we are seeking for, modulo a shift $m \rightarrow q-2$ in the coefficients (recall a remark after \eqref{Irr-q}).
In this way we obtain the exact formula for the generic coefficient:
\begin{equation}\label{coefficients_q_pos}
    c_k = \frac{1}{k!}\frac{(q-2 + k)!}{(q-2 + 2k)!}.
\end{equation}
Plugging it in into (\ref{phi_expansion}) results in the closed form of the superisospin expansion  for $q\ge 2$:
\begin{equation}
{\Phi}^{+q} = \hat{\Phi}^{+ q}_{I=\frac{q}{2}-1} + \sum_{k=1}^{\infty} \frac{(q-2+k)!}{(q-2+2k)!}\frac{1}{k!}(\nabla^{--})^k \hat{\Phi}^{+(q+2k)}_{I=\frac{q}{2}+k-1}.
\end{equation}

All this has been performed  in the Appendix A for the standard $SU(2)$ isospin case, so we can make use of the results collected there. Since the operator playing the role of $\partial^0$ is now $\nabla^{0}$, the
only difference is a shift the harmonic external $U(1)$ charge by an universal  constant for all analytic superfields
\footnote{The same idea can be applied for the superisospin expansion of general non-analytic superfield: one just needs to split it into different parts corresponding to possible eigenvalues
of $\nabla^0 \in \{q-2, q-1, q, q+1, q+2\}.$}:
\begin{equation}\label{Projectors/nablavalue}
\nabla^{0} \Phi^{+q} = (q - 2)\Phi^{+q}.
\end{equation}

So, for analytic superfields with the charges $q\ge 2$ one can make use of the same reasoning as  in the Appendix A, when deriving there the projectors on the definite isospin. Since every result
there is valid \textit{algebraically}, and $\nabla$ operators
satisfy the same algebra modulo an universal shift of the harmonic $U(1)$ charge, the \textit{same results can be used here}. Below we give a few examples.
\vspace{0.2cm}

\textbf{Off-shell ${\cal N}=2$ Yang-Mills prepotential $V^{++}$}. It is expanded as follows, just repeating (modulo the shift of the harmonic charge) the  $SU(2)$ expansion (\ref{fandomDecom}):

\begin{equation}
    V^{++} = \hat{V}^{++}_{I=0} + \frac{1}{2!} (\nabla^{--})\hat{V}^{+4}_{I=1} + \frac{1}{4!} (\nabla^{--})^2\hat{V}^{+6}_{I=2} + \frac{1}{6!} (\nabla^{--})^3\hat{V}^{+8}_{I=3} ... ,\label{IsoV}
\end{equation}
where the definitions of $\hat{V}^{+2m}_{I=m-1}$ closely follow eqs. (\ref{A86}), (\ref{Repl}):
\begin{equation}
    \hat{V}^{+2m}_{I=m-1} = V^{+2m} + \sum_{k=1}^{\infty} \frac{(-1)^k(2m-1)!}{k! (k+2m-1)!} (\nabla^{--})^{k}(\nabla^{++})^{k} V^{+2m},
    \quad V^{+2m} \overset{def}{=} (\nabla^{++})^{m-1} V^{++}. \label{IsoV2}
\end{equation}

It can be checked {\it algebraically} that the irreducible components satisfy the constraint:
\begin{equation}\label{constraint}
\nabla^{++}\hat{V}^{+2m}_{I=m-1} = 0.
\end{equation}
The projectors in this case read
\begin{equation}
    \text{П}^{(q=2)}_{I=m-1} = \sum_{k=0}^{\infty}\frac{(2m-1)(-1)^k}{k!(k+2m-1)!}(\nabla^{--})^{k+m-1}(\nabla^{++})^{k+m-1}, \quad m=1,2,...\, .
\end{equation}

The basic feature of ${\cal N}=2$ Yang-Mills theory is the invariance under the gauge transformations with analytic superfield parameter $\lambda(\zeta)$,

\bea
\delta V^{++} = D^{++} \lambda + {\rm nonlinear \; terms}\,. \label{gaugeV}
\eea
This gauge freedom allows one to impose (off shell) the gauge
\be
\nabla^{++}{V}^{++} = D^{++}{V}^{++} = 0\,. \label{Gauge}
\ee
In this gauge, all superisospins in $V^{++}$, apart from the lowest component $\hat{V}^{+2 }_{I=0}$,
fully drop out, which means that the gauge potential $V^{++}$ carries off shell only the superisospin $I=0$. All other superisospins are gauge degrees of freedom.

Note that in the case of {\it massive} ${\cal N}=2$ super Yang-Mills (see, e.g., \cite{BuPle}), the decomposition \eqref{IsoV}, \eqref{IsoV2} is valid also on shell,
with the replacement $\Box \Rightarrow M^2 \neq 0$ (see....). In this case the condition \eqref{Gauge} follows from the equations of motion,
\be
M^2 D^{++}{V}^{++} = 0\,.  \label{CondMass}
\ee
So the massive ${\cal N}=2$ super Yang-Mills carries on shell only the superisospin $I=0$.
\vspace{0.2cm}

\textbf{Off-shell ${\cal N}=2$ tensor multiplet}. It is described by the off-shell constrained analytic superfield $P^{++}$,
\be
D^{++} P^{++} = 0\,.\label{Tensor}
\ee
This constraint leaves in $P^{++}$ only the superisospin $I=0$ off shell. No any gauge freedom is associated with $P^{++}$ and so it describes some matter scalar ${\cal N}=2$
multiplet. Some generalizations of $P^{++}$, with higher-order harmonic constraints leaving more superisospins off shell, were discussed in \cite{MostGen}.
In the massless case the irreducible representations are characterized on shell by the
super-isohelicity, which avoids the appearance of the singular factors $\sim \Box^{-1}$  (see section 5).\\

As was already mentioned, for the analytic superfields with charges $q < 2$ the effective charge is negative and therefore they should satisfy the constraint \eqref{contraint_2}
with the derivative $\nabla^{--}$. The relevant decompositions are given below.
\vspace{0.2cm}

\textbf{Off-shell $q^+$ hypermultiplet.}  This procedure for hypermultiplet $q^+$ goes as follows: because eigenvalue of $\nabla^0$ on hypermultiplet is equal to -1, the expansion has the form:

\begin{equation}\label{formula}
q^+ = b_0 \hat{q}^+_{I=1/2} + b_1 \nabla^{++} \hat{q}^{-}_{I=1/2} + b_2 (\nabla^{++})^2 \hat{q}^{-3}_{I=3/2} + ...\,,
\end{equation}
where each component, instead of (\ref{constraint}),  is constrained by the condition\footnote{This constraint is non-trivial even for $q^+$, i.e. $m=0$, since $\nabla^{--}$ involves
a non-trivial term $\sim \Box^{-1}\partial_{\alpha\dot\alpha}D^{-\alpha}\bar D^{-\dot\alpha}$ which, e.g., relates $q^+|_{\theta = 0}$ with the component
$\sim A^{-}_{\alpha\dot\alpha}\theta^{+\alpha}\bar{\theta}^{+\dot\alpha}$. Due to these peculiarities, \eqref{HyperOS} does not imply any dynamical equations for the component fields in $q^+$, in contrast
to the standard equation of motion for $q^+$, $D^{++}q^+ = 0$.}:
\begin{equation}
\nabla^{--}\hat{q}^{+(1-2m)}_{I=m+1/2} = 0\,, \qquad m= 0, 1, 2, ...\, . \label{HyperOS}
\end{equation}

The reasoning quite similar to the case of $\Phi^q\,,\; q\geq 2$, leads to the expansion

\begin{equation}\label{formula2}
q^+ = \hat{q}^+_{I=1/2} + \frac{2}{3!} \nabla^{++} \hat{q}^{-}_{I=3/2} + \frac{3}{5!} (\nabla^{++})^2 \hat{q}^{-3}_{I=5/2} + ...\,,
\end{equation}
\begin{equation}
\hat{q}^{+(1-2m)}_{I=m+1/2} = q^{+(1-2m)} + \sum_{k=1}^{\infty} \frac{(-1)^k(2m+2)!}{k! (k+2m+2)!} (\nabla^{++})^{k}(\nabla^{--})^{k} q^{+(1-2m)}, \;
q^{+(1-2m)} \overset{def}{=} (\nabla^{--})^{m} q^{+}\,, \lb{Irrq+}
\end{equation}
with the projectors being:

\begin{equation}
\text{П}^{(q=1)}_{I=m+1/2} = \sum_{k=0}^{\infty}\frac{2(m+1)^2(-1)^k}{k!(k+2m+2)!}(\nabla^{++})^{k+m}(\nabla^{--})^{k+m}, \quad m = 0,1,2,...\,.
\end{equation}
As in the previous cases, the constraints \eqref{HyperOS} are satisfied purely algebraically, i.e, without implying any dynamics for $q^+$.

Note that the complete expansion \eqref{formula2} can be written as the following series
\bea
q^+ = \sum_{k=0}\frac{k + 1}{(2k +1)!} (\nabla^{++})^k \hat{q}^{+(1-2k)}_{I = k +1/2}\,. \label{Decq+}
\eea
\vspace{0.2cm}

\textbf{Off-shell $\omega$ hypermultiplet.} In this case
\bea
&&\omega =  \hat{\omega}^0_{I = 1} + b'_1 \nabla^{++}\hat{\omega}^{--} _{I = 2}+ b'_2(\nabla^{++})^2\hat{\omega}^{-4}_{I = 3} + \ldots   + b'_m(\nabla^{++})^m\hat{\omega}^{-2m}_{I = m + 1} + \ldots\,, \label{DecOmega} \\
&& \nabla^{--}\hat{\omega}^{-2m}_{I = m +1} = 0\,,  \quad m=0, 1, \ldots\, ,\label{ConstrOmega}
\eea
where $\omega$ is an arbitrary analytic zero-charge superfield, while $\hat{\omega}^{-2m}_{I = m + 1}$ are its fixed superisospin components, i.e. the analytic superfields constrained by eq. \eqref{ConstrOmega}.
For these irreducible components we obtain the following representation
\bea
\hat{\omega}^{-2m}_{I = m + 1} = {\omega}^{-2m} + \sum_{k=1} (-1)^k \frac{(2m + 3)!}{k!( k + 2m + 3)!}(\nabla^{++})^k(\nabla^{--})^k {\omega}^{-2m}\,, \;{\omega}^{-2m} := (\nabla^{--})^m \omega\,. \label{OmIrr}
\eea
These irreducible superfields satisfy \eqref{ConstrOmega} {\it{algebraically}}, without implying any dynamics for $\omega$. For the expansion \eqref{DecOmega} one can find a closed formula similar to \eqref{Decq+}.
The coefficients in that case follow the sequence:
\be
b'_1 = \frac{1}{4}\,, \quad b'_2 = \frac{1}{60}\,, \quad b'_3 = \frac{4\cdot5}{8!}\,, \qquad b'_{m-1} = \frac{m(m+1)}{(2m)!}\,.
\ee
The decomposition \eqref{DecOmega} can then be written as an infinite sum, like \eqref{Decq+},
\bea
\omega  = \sum_{k=0}\frac{(k+1)(k+2)}{[2(k+1)]!}(\nabla^{++})^k\hat{\omega}^{-2k}_{I = k+1}\,, \quad C_0 = 1\,, \lb{DecOmega2}
\eea
with the  projectors being:
\begin{equation}
\text{П}^{(q=0)}_{I=1+m} = \sum_{k=0}^{\infty}\frac{(-1)^k(2m+3)(m+1)(m+2)}{k!(k+2m+3)!}(\nabla^{++})^{k+m}(\nabla^{--})^{k+m}, \quad m = 0,1,2,...\,.
\end{equation}

Finally, let us recall the expansion of a general analytic superfield $\Phi^{+q}$ with $q \geq 2$ over irreducible super-isospins
\begin{align}
& \Phi^{+q} = \hat{\Phi}^{+q}_{I = q/2-1} + \sum_{k=1}\frac{1}{k!}\frac{(q-2 + k)!}{(q-2 + 2k)!}(\nabla^{--})^k\hat{\Phi}^{+(q + 2k)}_{I = q/2+k-1}\,, \lb{StandPhiq} \\
& \nabla^{++}\hat{\Phi}^{+(q + 2m)}_{I = q/2+m-1} = D^{++}\hat{\Phi}^{+(q + 2m)}_{I = q/2+m-1} = 0\,, \lb{StandPhiqCon}\\
& \hat{\Phi}^{+(q + 2m)}_{I = q/2+m-1} =\Phi^{+(q + 2m)} +
\sum_{k=1}(-1)^k\frac{(q + 2m -1)!}{k!(k + q +2m -1)!}(\nabla^{--})^k (\nabla^{++})^k\Phi^{+(q + 2m)}\,, \label{StandPhiq2} \\
& \text{П}^{(q)}_{(I=q/2 + m - 1)} = \frac{1}{m!}\frac{(q-2+m)!}{(q-2+2m)!}\sum_{k=0}\frac{(-1)^k (q+2m-1)!}{k!(k+q+2m-1)!}(\nabla^{--})^{k+m}(\nabla^{++})^{k+m}\,.
\end{align}
We observe that the superfields which are irreducible with respect to the given superisospin appear both in the expansions of the type \eqref{StandPhiq} and in those \eqref{Decq+}, \eqref{DecOmega2}. Obviously, they should
coincide modulo a numerical coefficient. The proportionality coefficients can be explicitly found, but generally this is a rather cumbersome procedure.
For the lowest values of the superisospin it is much simpler.
For instance, consider the first irreducible component with $q=3, k=0$  in \eqref{StandPhiq},
\bea
\hat{h}^{+3}_{I = 1/2} = h^{+3} +\sum_{k=1}(-1)^k\frac{2}{k!(k + 2)!}(\nabla^{--})^k (\nabla^{++})^k h^{+ 3}\,.
\eea
Choosing, without loss of generality,
$$
h^{+3} = \nabla^{++}q^{+}\,,
$$
it is straightforward to prove that
\be
\hat{h}^{+3}_{I = 1/2} = \nabla^{++}\hat{q}^+_{I = 1/2} = D^{++}\hat{q}^+_{I = 1/2}\,, \lb{Relhq}
\ee
where, in accord with the general formula \eqref{Irrq+},
\be
\hat{q}^+_{I = 1/2} = q^{+} + \sum_{k=1}^{\infty} \frac{2 (-1)^k}{k! (k +2)!} (\nabla^{++})^{k}(\nabla^{--})^{k} q^{+}\,. \lb{q+Irr}
\ee
The covariant derivative $\nabla^{++}$ commutes with the superisospin Casimir, so the objects in the left and right sides of \eqref{Relhq} carry
the same superisospin\footnote{The proof of \eqref{q+Irr} is based on the simple identity $[(\nabla^{++})^n, (\nabla^{--})^n ]q^+ = n^2(\nabla^{--})^{n-1} (\nabla^{++})^{n-1} q^+$.}.

\section{Massless case}

In the free massless  case, one has on shell:
\be
\Box \Phi = -P^2 \Phi = 0\,. \label{Box}
\ee
This equation imposes strong constraints on supersymmetry generators and covariant derivatives.

Proceeding from the identity:
\begin{equation}\label{Casimir/Generators/Massless/zeroanticommutator}
    \{P^{\dot{\alpha}\alpha}Q_{\alpha}^i, P^{\dot{\beta}\beta}\bar{Q}_{j\dot{\beta}}\} = -\Box P^{\dot{\alpha}\beta}\delta^i_j,
\end{equation}
one can derive, like in $\mathcal{N}=1$ case \cite{Buchbinder:1998twe,ArvMezTown}, the following generic Dirac-like constraints for the supercharges and spinor covariant derivatives,
\begin{align}\label{Casimir/Generators/Massless/Qconstraints}
    P_{\alpha\dot{\alpha}}Q^{\alpha}_i \Phi = P_{\alpha\dot{\alpha}}\bar{Q}^{\dot{\alpha}}_i \Phi = 0, \quad
    P_{\alpha\dot{\alpha}}D^{\alpha}_i \Phi = P_{\alpha\dot{\alpha}}\bar{D}^{\dot{\alpha}}_i \Phi = 0.
\end{align}
The second pair of equations follows from the first one, taking into account the relation between $Q$ and $D$,
\bea \lb{RelQD}
Q^i_\alpha = iD^i_\alpha + 4\bar{\theta}^{i\dot\alpha}\partial_{\alpha\dot\alpha}\,, \quad \bar{Q}_{i\dot\alpha} = i\bar{D}_{i\dot\alpha} -4{\theta}^\alpha_{i}\partial_{\alpha\dot\alpha}\,,
\eea
together with the zero-mass equation $\Box \Phi = 0$.

Below (in section 6) we will show that in the massless case eqs. \eqref{Box} and \eqref{Casimir/Generators/Massless/Qconstraints} are satisfied on shell
for some gauge invariant objects $\Phi$ in most ${\cal N}=2$ theories of interest. Notice that on the surface of eqs. \eqref{Casimir/Generators/Massless/Qconstraints}
 the following commutation relations also hold
\bea \label{Casimir/Generators/Massless/zerocommutator}
[Q^{\alpha k}Q_\alpha^i, \bar{Q}_{j\dot\beta}] = [Q^{\alpha k}Q_\alpha^i, Q^j_\beta] = 0 \;\;({\rm and \; c.c.}),
\eea
which are compatible with extra on-shell free field constraints
\bea \label{Casimir/Generators/Massless/Qconstraints2}
(Q^{\alpha k}Q_\alpha^i)\Phi \simeq (D^{\alpha k}D_\alpha^i)\Phi \simeq 0 \;\;({\rm and \; c.c.}).
\eea
They also turn out to be valid in the theories discussed in section 6.

\subsection{Superhelicity}
To determine irreducible representations in the massless case, one needs to define the proper  super extension of helicity, the superhelicity operator.
We need to modify Pauli-Lubanski vector in a  way which avoids the appearance  of the singular operator $\Box^{-1}$ and still preserves the commutativity
with supercharges. Like in the case of the superspin, there are two different operators appropriate for this role.
One of them uses the $R$ symmetry generator \cite{Gates} and another does not \cite{Buchbinder:1998twe}. We stick to  the first option:
\begin{align}
    \mathbb{W}_{\alpha\dot\alpha} = W_{\alpha\dot\alpha} + \frac{1}{8}[Q^i_{\alpha},\bar{Q}_{i\dot{\alpha}}]  - \frac{i}{2} R P_{\alpha\dot\alpha}, \lb{SupHel}
\end{align}
since it commutes with supercharges $Q$ off shell, without exploiting Dirac-like equations (\ref{Casimir/Generators/Massless/Qconstraints}):
\begin{equation}
    [\mathbb{W}_{\alpha\dot\alpha}, Q^i_{\beta}] = [\mathbb{W}_{\alpha\dot\alpha}, \bar{Q}_{i\dot{\beta}}] = 0\,. \lb{QW}
\end{equation}
This commutativity property can be checked using the relations \eqref{CommWQ} and the last relation in \eqref{RTrel}.

In the analytic basis this operator reads:
\begin{equation}\label{superhelicity_analytic}
        \mathbb{W}_{\alpha\dot{\alpha}} =
        \mathcal{W}_{\alpha\dot{\alpha}}
        +\frac{1}{4}(D^-_{\alpha}\bar{D}^{+}_{\dot{\alpha}} + \bar{D}^-_{\dot{\alpha}}D^+_\alpha)
        -\frac{1}{2}\mathcal{R}\partial_{\alpha\dot{\alpha}}\,.
\end{equation}
Now we pass on shell, {\it i.e.} use eqs. (\ref{Casimir/Generators/Massless/Qconstraints}). We have
\begin{equation}\label{Psqr=0,WP=0}
    P^2 = 0, \quad \mathbb{W}^{\dot{\alpha}\alpha}P_{\alpha\dot{\alpha}} \simeq 0
\end{equation}
(note that $\mathcal{W}_{\alpha\dot{\alpha}}P^{\alpha\dot\alpha} \sim \Box \simeq 0$ in virtue of the definition \eqref{CommModW}). The square of $\mathbb{W}$ can be either zero or non-zero.
In the latter case one encounters continuous spectrum,
and we do not consider this option. So we assume
\begin{equation}
    \mathbb{W}^{\dot{\alpha}\alpha}\mathbb{W}_{\alpha\dot{\alpha}} = 0. \lb{W2zero}
\end{equation}
For any two vectors $A_{\alpha\dot{\alpha}}$ and $B_{\alpha\dot{\alpha}}$ with zero norm  the following property holds:
\begin{equation}\label{vector_statement}
    A^2=0, \quad B^2=0, \quad A\cdot B = 0, \ \Rightarrow \ A = \lambda B\,.
\end{equation}
Combining \eqref{W2zero} with (\ref{Psqr=0,WP=0} and applying the property just mentioned, we find
\begin{equation}
    \mathbb{W}_{\alpha\dot{\alpha}} \simeq \lambda^{H} P_{\alpha\dot\alpha} = -i\lambda^{H} \partial_{\alpha\dot{\alpha}}\,,
\end{equation}
where $\lambda^H$ is called superhelicity. For superfields with external Lorentz indices
$\Phi^{+q}_{\alpha_1...\alpha_n\dot{\alpha}_1...\dot{\alpha}_m}$  some extra conditions are needed like in the $\mathcal{N}=1$ case
\cite{Buchbinder:1998twe}. For sake of simplicity we do not consider such complications here.

The process of recovering the helicity contents of massless ${\cal N}=2$ supermultiplets with a given superhelicity will be exemplified below. It is based on the commutators
\bea \lb{DDW}
[D^{\pm}_\beta, \mathbb{W}_{\alpha\dot\alpha}] = iD^{\pm}_\alpha \partial_{\beta\dot\alpha} \simeq iD^{\pm}_\beta \partial_{\alpha\dot\alpha}\,, \quad
[\bar{D}^{\pm}_{\dot\beta}, \mathbb{W}_{\alpha\dot\alpha}] = -i\bar{D}^{\pm}_{\dot\alpha} \partial_{\alpha\dot\beta} \simeq -i\bar{D}^{\pm}_{\dot\beta} \partial_{\alpha\dot\alpha}\,.
\eea

\subsection{Super-isohelicity}
Extending the notion of superisospin to the massless case was given in \cite{Gates}. The basic object in this case reads:
\begin{align}
    \mathbb{W}^{i}_{j\alpha\dot{\alpha}} &= T^i_j\partial_{\alpha\dot{\alpha}}
    -\frac{1}{8}\left([Q^i_\alpha, \bar{Q}_{j\dot{\alpha}}]-\frac{1}{2}\delta^i_j [Q^k_\alpha, \bar{Q}_{k\dot{\alpha}}]\right) \label{Wops_definition}.
\end{align}
We should check under which conditions  these operators commute with supercharges.
Consider commutators of the two parts of (\ref{Wops_definition}) with supercharges $Q^{k}_{\hat\beta} = (Q^{k}_{\beta}, \,\bar{Q}^{k}_{\dot\beta})$:
\begin{align}
    [T^i_j \partial_{\alpha\dot{\alpha}}, Q^k_{\hat\beta}] &= -i\partial_{\alpha\dot{\alpha}}\left(\delta^k_j Q^i_{\hat\beta} - \frac{1}{2}\delta^i_j Q^k_{\hat\beta} \right), \label{Wcommutators_1}  \\
    \frac{1}{8}\left[[Q^i_\alpha, \bar{Q}_{j\dot{\alpha}}]-\frac{1}{2}\delta^i_j [Q^l_\alpha, \bar{Q}_{l\dot{\alpha}}], Q^k_{\beta} \right]&
    = -i\partial_{\beta\dot{\alpha}}\left(\delta^k_j Q^i_{\alpha} - \frac{1}{2}\delta^i_j Q^k_{\alpha}\right), \label{Wcommutators_2} \\
\frac{1}{8}\left[[Q^i_\alpha, \bar{Q}_{j\dot{\alpha}}]-\frac{1}{2}\delta^i_j [Q^l_\alpha, \bar{Q}_{l\dot{\alpha}}], \bar{Q}^k_{\dot\beta} \right]&
    = -i\partial_{\alpha\dot{\beta}}\left(\delta^k_j \bar{Q}^i_{\dot\alpha} - \frac{1}{2}\delta^i_j \bar{Q}^k_{\dot\alpha}\right).\label{Wcommutators_3}
\end{align}
We observe that on the right hand sides of (\ref{Wcommutators_2}) and (\ref{Wcommutators_3}) the indices $\alpha, \beta$ (or $\dot\alpha, \dot\beta$) are switched compared to (\ref{Wcommutators_1}). So the difference of
(\ref{Wcommutators_2}), (\ref{Wcommutators_3}) and (\ref{Wcommutators_1}) is proportional to the on-shell Dirac-type equations (\ref{Casimir/Generators/Massless/Qconstraints}), which means
that on shell we do indeed have:
\begin{equation}
    [\mathbb{W}^{i}_{j\alpha\dot{\alpha}}, Q^{k}_{\hat{\beta}}] \simeq 0\,.
\end{equation}
This makes the massless case radically different from the massive one, since the Dirac-like equations are needed to secure the commutativity of \eqref{Wops_definition}
with ${\cal N}=2$ supercharges. This resort to the on-shell equations is in fact the only way to avoid the appearance of the singular operator $\Box^{-1}$.

The operators (\ref{Wops_definition}) can be rewritten in terms of spinor covariant derivatives as
\bea
\mathbb{W}^{(ij)}_{\alpha\dot{\alpha}} = T^{(u)(ij)}\partial_{\alpha\dot{\alpha}}  + \frac{1}{4} D^{(i}_\alpha\bar{D}^{j)}_{\dot\alpha}\,, \lb{ThroughDD}
\eea
where we applied to the relations \eqref{RelQD} and used once more the dynamical equation (\ref{Casimir/Generators/Massless/Qconstraints}). In analytic basis,
the operators (\ref{Wops_definition})
are projected onto harmonics according to:
\begin{equation} \lb{WProj}
    \mathbb{W}^{ij}_{\alpha\dot{\alpha}} = iu^{+i}u^{+j}\mathbb{W}^{--}_{\alpha\dot{\alpha}} - iu^{-i}u^{-j}\mathbb{W}^{++}_{\alpha\dot{\alpha}} - iu^{+(i}u^{-j)}\mathbb{W}^{0}_{\alpha\dot{\alpha}}\,.
\end{equation}

Projecting the right-hand side of \eqref{ThroughDD} on the harmonics with he help of relations \eqref{TuExpan}  and \eqref{def/spincovderanalytic}, we finally obtain
\begin{align}
    \label{Wpmpmexpression}
\mathbb{W}^{\pm\pm}_{\alpha\dot{\alpha}} &\simeq D^{\pm\pm}\partial_{\alpha\dot{\alpha}} \pm \frac{i}{4}D^\pm_\alpha \bar{D}^\pm_{\dot{\alpha}}\,,\\
\label{W0expression}
\mathbb{W}^{0}_{\alpha\dot{\alpha}} &\simeq D^0 \partial_{\alpha\dot{\alpha}} -  \frac{i}{4}\left(D^+_\alpha \bar{D}^-_{\dot{\alpha}} + D^-_\alpha \bar{D}^+_{\dot{\alpha}}\right)=
(D^0 -1)\partial_{\alpha\dot{\alpha}} + \frac{i}{4}\left(\bar{D}^-_{\dot{\alpha}}D^+_\alpha  - D^-_\alpha \bar{D}^+_{\dot{\alpha}}\right).
\end{align}
Like in the off-shell (or massive on-shell) cases, in the on-shell massless case under consideration we once again observe the shift of the harmonic $U(1)$ charge on the analytic superfields,
this time by amount $(-1)$.

The non-vanishing commutators of these operators  with spinor covariant derivatives are as follows:
\begin{align}\label{Woperatorscommutators}
    [ D^\pm_{\gamma}, \mathbb{W}^{0}_{\alpha\dot{\alpha}}] &= \mp\epsilon_{\gamma\alpha}D^{\pm\gamma'} \partial_{\gamma'\dot{\alpha}}, &
    [\bar{D}^\pm_{\dot{\gamma}}, \mathbb{W}^{0}_{\alpha\dot{\alpha}}] &= \mp\epsilon_{\dot{\gamma}\dot{\alpha}}\bar{D}^{\pm\dot{\gamma}'} \partial_{\alpha\dot{\gamma}'} \\ \label{Woperatorscommutators2}
    [ D^\mp_{\gamma}, \mathbb{W}^{\pm\pm}_{\alpha\dot{\alpha}}] &= -\epsilon_{\gamma\alpha}D^{\pm\gamma'} \partial_{\gamma'\dot{\alpha}}, &
    [\bar{D}^\mp_{\dot{\gamma}}, \mathbb{W}^{\pm\pm}_{\alpha\dot{\alpha}}] &= -\epsilon_{\dot{\gamma}\dot{\alpha}}\bar{D}^{\pm\dot{\gamma}'} \partial_{\alpha\dot{\gamma}'}.
\end{align}
On the surface of Dirac-like equations (\ref{Casimir/Generators/Massless/Qconstraints}) all these commutators vanish:
\begin{equation}\label{Wops_spinor_commutators}
    [\mathbb{W}^{0}_{\alpha\dot{\alpha}}, D^{\pm}_{\hat{\beta}}] \simeq     [\mathbb{W}^{\pm\pm}_{\alpha\dot{\alpha}}, D^{\mp}_{\hat{\beta}}] \simeq 0\,.
\end{equation}
Also, it is useful to give non-vanishing commutators with harmonic derivatives
\bea
[D^{\pm\pm}, \mathbb{W}^{\mp\mp}_{\alpha\dot\alpha}] = \pm \mathbb{W}^0_{\alpha\dot\alpha}\,, \; [D^0, \mathbb{W}^{\pm\pm}_{\alpha\dot\alpha}] = \pm 2\mathbb{W}^{\pm\pm} _{\alpha\dot\alpha}\,, \;
[D^{\pm\pm}, \mathbb{W}^{0}_{\alpha\dot\alpha}] = \mp 2 \mathbb{W}^{\pm\pm}_{\alpha\dot\alpha}\,. \label{WithHarmDer}
\eea

Obviously, the objects $\mathbb{W}^{\pm\pm}_{\alpha\dot{\alpha}}, \mathbb{W}^{0}_{\alpha\dot{\alpha}}$ are the massless-case analogs
of the superisospin constituents $\nabla^{\pm\pm}, \nabla^0$.  To better understand the algebraic on-shell structure of the massless ${\cal N}=2$ theories, one
needs to know the commutation relations  between these operators. Once again, we are interested in them only modulo Dirac-like and Box constraints,
as before. We obtain
\begin{subequations}\label{Wops_commutation}
    \begin{align}
        [\mathbb{W}^{++}_{\alpha\dot{\alpha}}, \mathbb{W}^{--}_{\beta\dot{\beta}}] &\simeq
        \mathbb{W}^0_{\beta\dot{\beta}}\partial_{\alpha\dot{\alpha}} \simeq \mathbb{W}^0_{\alpha\dot{\alpha}}\partial_{\beta\dot{\beta}} \simeq
        \mathbb{W}^0_{(\alpha(\dot{\alpha}}\partial_{\beta)\dot{\beta})}\,, \label{WppWmm} \\
        [\mathbb{W}^{0}_{\alpha\dot{\alpha}}, \mathbb{W}^{\pm\pm}_{\beta\dot{\beta}}] &\simeq
         \pm 2\mathbb{W}^{\pm\pm}_{\beta\dot{\beta}}\partial_{\alpha\dot{\alpha}} \simeq
         \pm 2\mathbb{W}^{\pm\pm}_{\alpha\dot{\alpha}}\partial_{\beta\dot{\beta}} \simeq \pm
         2\mathbb{W}^{\pm\pm}_{(\alpha(\dot{\alpha}}\partial_{\beta)\dot{\beta})}\,, \label{W0Wpmpm}\\
 [\mathbb{W}^{0}_{\alpha\dot{\alpha}}, \mathbb{W}^{0}_{\beta\dot{\beta}}] &\simeq 0\,, \,[\mathbb{W}^{\pm\pm}_{\alpha\dot{\alpha}}, \mathbb{W}^{\pm\pm}_{\beta\dot{\beta}}] \simeq 0\,,
[\mathcal{\mathbb{W}}_{\alpha\dot{\alpha}}, \mathbb{W}^{i}_{j\beta\dot{\beta}}] = 0\,.
    \end{align}
\end{subequations}
These commutation relations resemble those of $su(2)$ algebra, with $\mathbb{W}^{0}_{\alpha\dot{\alpha}}$ playing the role of $U(1)$ ``charge'',
and $\mathbb{W}^{\pm\pm}_{\alpha\dot{\alpha}}$ of raising and lowering operators. Despite the unavoidable presence of spinorial Lorentz indices in \eqref{Wops_commutation}\footnote{In fact,
the generators $\mathbb{W}^{\pm\pm}_{\alpha\dot{\alpha}}, \mathbb{W}^{0}_{\alpha\dot{\alpha}}$ taken together with the right-hand sides of the commutators \eqref{Wops_commutation}
form an infinite-dimensional non-abelian algebra with the generic generators of the type $\mathbb{W}^{\pm\pm}_{(\alpha(\dot{\alpha}}\partial_{\alpha_1\dot\alpha_1}\cdots
\partial_{\alpha_n)\dot\alpha_n)}$, where the total symmetrization is assumed separately in the undotted and dotted indices and eqs. \eqref{Box},
\eqref{Casimir/Generators/Massless/Qconstraints} are supposed to be valid. Any commutator in such an algebra is entirely specified by those in \eqref{Wops_commutation} due to commutativity of
4- translation generators.}
one could try to construct a representation of this algebra on the on-shell ${\cal N}=2$ superfields following the analogy with $su(2)$ just mentioned. Namely, one could define
the  ``highest weight'' state $\Phi^{+q}$ by imposing
\bea
(a)\;\, \mathbb{W}^{++}_{\alpha\dot{\alpha}}\Phi^{+q} = 0 \quad {\rm or} \quad (b)\;\, \mathbb{W}^{--}_{\alpha\dot{\alpha}}\Phi^{+q} = 0\,,
\eea
and construct an irreducible ``module'', successively acting on $\Phi^{+q}$ by $\mathbb{W}^{--}_{\alpha\dot{\alpha}}$, or $\mathbb{W}^{++}_{\alpha\dot{\alpha}}$, respectively,
and using the commutation relations \eqref{Wops_commutation}.
Due to the commutativity conditions \eqref{Wops_spinor_commutators}, all superfields obtained in this way satisfy both Box
and Dirac-like equations \eqref{Box} and \eqref{Casimir/Generators/Massless/Qconstraints},
like $\Phi^{+q}$ itself.

Furthermore, we can define an analog of the Casimir operator of $su(2)$,
\begin{equation}\label{superisohelicity_operator}
    C^{[\mathcal{I}]}_{\alpha\beta\dot{\alpha}\dot{\beta}} = \frac{1}{4}\mathbb{W}^{0}_{\alpha\dot{\alpha}} \mathbb{W}^{0}_{\beta\dot{\beta}}
    + \frac{1}{2}\mathbb{W}^{++}_{\alpha\dot{\alpha}} \mathbb{W}^{--}_{\beta\dot{\beta}} +\frac{1}{2}\mathbb{W}^{--}_{\alpha\dot{\alpha}} \mathbb{W}^{++}_{\beta\dot{\beta}}\,.
\end{equation}
Using (\ref{Wops_commutation}), one can check that it commutes with the whole superalgebra
$\mathbb{W}^{\pm\pm}_{\alpha\dot{\alpha}}, \mathbb{W}^{0}_{\alpha\dot{\alpha}}$ (again modulo dynamical constraints). It also commutes with spinorial and harmonic derivatives.  Defining the ``vacuum state'' by the conditions
\bea \lb{W++Cond}
\mathbb{W}^{++}_{\alpha\dot{\alpha}}\Phi^{+q} = 0\,, \label{++0}
\eea
we find
\bea
C^{[\mathcal{I}]}_{\alpha\beta\dot{\alpha}\dot{\beta}}\Phi^{+q} = \frac{1}{4}\Big(\mathbb{W}^{0}_{\alpha\dot{\alpha}} \mathbb{W}^{0}_{\beta\dot{\beta}}
+ \mathbb{W}^{0}_{\beta\dot{\beta}}\partial_{\alpha\dot{\alpha}} +\mathbb{W}^{0}_{\alpha\dot{\alpha}}\partial_{\beta\dot{\beta}}\Big)\Phi^{+q}\,.
\eea

Analogously, for the choice
\bea
\mathbb{W}^{--}_{\alpha\dot{\alpha}}\hat{\Phi}^{+q} = 0\,, \lb{--0}
\eea
we have
\bea
\hat{C}^{[\mathcal{I}]}_{\alpha\beta\dot{\alpha}\dot{\beta}}\hat{\Phi}^{+q} = \frac{1}{4}\Big(\mathbb{W}^{0}_{\alpha\dot{\alpha}} \mathbb{W}^{0}_{\beta\dot{\beta}}
- \mathbb{W}^{0}_{\beta\dot{\beta}}\partial_{\alpha\dot{\alpha}} -\mathbb{W}^{0}_{\alpha\dot{\alpha}}\partial_{\beta\dot{\beta}}\Big)\hat{\Phi}^{+q}\,.
\eea

For fixing the eigenvalue of $\mathbb{W}^{0}_{\beta\dot{\beta}}$ we can apply to the same arguments as in the case of super-helicity (eqs. \eqref{vector_statement}).
Indeed, starting from \eqref{W0expression}  and using both Box and Dirac-like equations, we can deduce
\bea
(\mathbb{W}^{0})^{\dot{\alpha}\alpha}(\mathbb{W}^{0})_{\alpha\dot{\alpha}} &\simeq -\frac{1}{16}[(D^-)^2(\bar{D}^+)^2 + (\bar{D}^{-})^2(D^+)^2 + 2\cdot  (D^-D^+)(\bar{D}^-\bar{D}^+)]\,.
\eea
For  most cases of interest (see section 6) the r.h.s. of this expression vanishes, which is evident, e.g.,  when it acts on the analytic superfields. So we have
\bea
\mathbb{W}^{0\,\alpha\dot{\alpha}}\mathbb{W}^{0}_{\alpha\dot{\alpha}} \simeq 0, \quad \mathbb{W}^{0\,\alpha\dot{\alpha}}P_{\alpha\dot{\alpha}}\simeq 0\,,\quad P^2 = 0\,,
\eea
whence it follows (as in \eqref{vector_statement}) that
\be \lb{W0Gen}
\mathbb{W}_{0\,\alpha\dot{\alpha}} \simeq g_0 \partial_{\alpha\dot{\alpha}}\,.
\ee
By analogy with the definition of helicity in \eqref{vector_statement}, the number $g_0 $ can be naturally dubbed ``super-isohelicity''.

In both cases, \eqref{++0} and \eqref{--0}, we have
\bea
&& {C}^{[\mathcal{I}]}_{\alpha\beta\dot{\alpha}\dot{\beta}} = \chi\,\partial_{(\alpha(\dot\alpha} \partial_{\beta)\dot\beta)}\,, \quad
\hat{C}^{[\mathcal{I}]}_{\alpha\beta\dot{\alpha}\dot{\beta}} = \hat\chi\,\partial_{(\alpha(\dot\alpha} \partial_{\beta)\dot\beta)}\,,\lb{Spectra} \\
&&\chi = \frac{g_0}{2}\Big( \frac{g_0}{2} + 1\Big), \quad \hat\chi = \frac{g_0}{2}\Big( \frac{g_0}{2} - 1\Big). \lb{Spectrag0}
\eea

Actually, for any massless model where the on-shell constraints \eqref{Casimir/Generators/Massless/Qconstraints}, \eqref{Casimir/Generators/Massless/Qconstraints2} are valid one can show that
the operator ${C}^{[\mathcal{I}]}_{\alpha\beta\dot{\alpha}\dot{\beta}}$ is reduced to its fully symmetric part
\bea
C^{[\mathcal{I}]}_{\alpha\beta\dot{\alpha}\dot{\beta}} \simeq \mathcal{I}(\mathcal{I}\pm 1)\partial_{(\alpha(\dot\alpha} \partial_{\beta)\dot\beta)}\,,
\eea
where the sign refers to two possible conditions \eqref{++0} and \eqref{--0}. This sign and the value of $\mathcal{I}$ fully characterize the isospin
contents of the given on-shell harmonic superfield $\Phi^{+q}$.

For determining the precise isospin contents of $\Phi^{+q}$, one needs to expand it over $\theta$ coordinates and for each $u$-dependent component in this expansion
to calculate the value of the covariant $SU(2)$ Casimir operator $\tilde{C}_I$  defined according to  (\ref{isospin_operator}). As was already noticed, the whole action of the group
$SU(2)_A$ in the analytic basis is reduced to its action on harmonics. It is curious that the operator $\tilde{C}_I$ can be efficiently used to specify the isospin contents
of the component fields in the on-shell analytic superfields if one defines these fields as the proper superfield projections and employs the commutation relations \eqref{HarSpin}.

\subsection{Super U(1)-helicity}

By the same arguments as in the case of superhelicity (\ref{Wops_definition}) one can construct the quantity commuting on shell with the supercharges
\begin{align}
    \mathbb{W}^{U(1)}_{\alpha\dot{\alpha}} &= R \partial_{\alpha\dot{\alpha}}-\frac{1}{8}[Q^k_\alpha, \bar{Q}_{k\dot{\alpha}}]\,.
\end{align}
In the analytic basis it reads:
\begin{align}
    \mathbb{W}^{U(1)}_{\alpha\dot{\alpha}} &\simeq \mathcal{R}\partial_{\alpha\dot{\alpha}}
    - \frac{1}{4}(D^-_{\alpha} \bar{D}^+_{\dot{\alpha}} + \bar{D}^-_{\dot{\alpha}}D^+_{\alpha}). \lb{Rsymmomega}
\end{align}
Its commutation relations with spinor covariant derivatives are:
\begin{align}
    [\mathbb{W}^{U(1)}_{\alpha\dot{\alpha}}, D^{\pm}_{\gamma}] &= i D^{\pm}_{\alpha}\partial_{\gamma\dot{\alpha}}, \\
    [\mathbb{W}^{U(1)}_{\alpha\dot{\alpha}}, \bar{D}^{\pm}_{\dot{\gamma}}] &= -i\bar{D}^{\pm}_{\dot{\alpha}}\partial_{\alpha\dot{\gamma}}.
\end{align}

Following the same logic as in the case of superhelicity, we should have on shell:
\begin{equation}
    \mathbb{W}^{U(1)}_{\alpha\dot{\alpha}} \simeq \lambda^{U(1)} P_{\alpha\dot{\alpha}} = -i \lambda^{U(1)} \partial_{\alpha\dot{\alpha}}\,,\lb{U1hel}
\end{equation}
where $\lambda^{U(1)}$ can be called \textit{super-$U(1)$-helicity}.

\section{Examples of massless free systems}

\subsection{ $q^{+}$ hypermultiplet}

The hypermultiplet is described by an analytic (\ref{analyticsuperfield}) superfield $q^{+}(\zeta)$
subject to the following two equivalent forms of the equation of motion:
\begin{equation}\label{qEqs}
(a)\;    D^{++}q^{+} = 0, \quad \leftrightarrow \quad (b)\; (D^{--})^2 q^{+} = 0\,.
\end{equation}
The equivalence of eqs. $(a)$ and $(b)$ can be proved using the {\it Lemma} \cite{Book}:
\be
D^{++}\Phi^{q} = 0 \quad \rightarrow \quad \Phi^{q} = 0 \;\; {\rm for} \; q < 0\,.\lb{Lemma1}
\ee
Though these two forms are equivalent at the free-field level, the first form is distinguished in that it can be derived
from the off-shell action
$$
\sim d\zeta^{(-4)} \tilde{q}^+ D^{++} q^+ + {\rm c.c.}\,, \qquad d\zeta^{(-4)} = d^4 x_A d^2\theta^+ d^2\bar\theta^+ du\,,
$$
which admits a generalization to the case with interaction \cite{Book}.

Hitting the second form  of the equation \eqref{qEqs} with $D^+_{\hat{\alpha}}$
and then making use of the first form together with the Grassmann analyticity condition $D^+_{\hat\alpha} q^+ = 0$, one obtains:
\begin{equation}\label{qplus_equations1}
    D^{--}{D}^{-}_{\hat{\alpha}} q^{+} = 0\,.
\end{equation}
Applying $D^+_{\hat{\alpha}}$ to \eqref{qplus_equations1} gives the equations:
\begin{align}\label{qplus_equations2}
    &(D^-)^2 q^{+} = 0, \quad (\bar{D}^-)^2 q^{+} = 0, \\
    &\big(D^{--}\partial_{\alpha\dot{\alpha}} - \frac{i}{4} D^{-}_{\alpha}\bar{D}^{-}_{\hat{\alpha}}\big)q^+ = 0\,. \label{qplus_equations3}
\end{align}
Hitting  \eqref{qplus_equations2} by $D^+_{\hat{\alpha}}$ one more time yields just the Dirac equations :
\begin{equation}\label{qplus_equations4}
    \partial_{\alpha\dot{\alpha}}{D}^{-{\alpha}} q^{+} = \partial_{\alpha\dot{\alpha}}\bar{D}^{-\dot{\alpha}} q^{+} = 0\,.
\end{equation}
Continuing the process, we come to the equation
\begin{equation}
    \Box q^{+} = 0. \lb{Boxq}
\end{equation}
Note that hitting of \eqref{qplus_equations3} by $D^+_{\hat{\alpha}}$  again yields \eqref{qplus_equations4}.
Specializing, e.g., to $D^+_{\alpha}$, we obtain
\bea
\partial_{\alpha\dot\alpha} D^-_\beta q^+ - \partial_{\beta\dot\alpha}D^-_{\alpha}q^+ \;\propto \; \varepsilon_{\alpha\beta}\partial_{\gamma\dot\alpha}D^{-\gamma}q^+ = 0.  \nonumber
\eea
Further action of the spinor and harmonic derivatives on the equations obtained does not provide any new information and serves just to express some higher-order component fields as
$x$ derivatives of the physical fields $q^+|_{\theta = 0}, (D^-_{\hat\alpha}q^+)|_{\theta=0}$ and to eliminate infinite tails of the auxiliary
fields appearing in the harmonic expansions.

One comment is necessary concerning the Dirac equations \eqref{qplus_equations4}. Being combined with the purely kinematical off-shell conditions following from the analyticity of $q^+$,
\bea
\partial_{\alpha\dot{\alpha}}{D}^{+{\alpha}} q^{+} = \partial_{\alpha\dot{\alpha}}\bar{D}^{+\dot{\alpha}} q^{+} = 0,
\eea
they amount to the free Dirac equations in their generic original central-basis form \eqref{Casimir/Generators/Massless/Qconstraints}. Also, eqs. \eqref{qplus_equations2} and those
obtained through action of  $D^{++}$ on them,
are combined into the general admissible conditions  \eqref{Casimir/Generators/Massless/Qconstraints2}.

Looking at eq. \eqref{qplus_equations3}, we observe that it is precisely the ``highest weight'' condition \eqref{--0},
$$
\mathbb{W}^{--}_{\alpha\dot\alpha} q^+ = 0\,.
$$
Also, it is direct to check that
$$
\mathbb{W}^{++}_{\alpha\dot\alpha} q^+ = \mathbb{W}^{0}_{\alpha\dot\alpha} q^+ = 0\,.
$$
Thus, on shell $q^+$ is a singlet of the whole algebra \eqref{Wops_commutation}.
Correspondingly, the ``super-isohelicity'' $g_0$ defined in \eqref{W0Gen}, equally as the Casimir operator
${C}^{[\mathcal{I}]}_{\alpha\beta\dot{\alpha}\dot{\beta}}$, take zero values on $q^+$:
\be
g^{(q)}_0 = \mathcal{I}^{(q)} = 0\,.
\ee


\textbf{Multiplet helicity content.} Since $q^+$ and its tilde-conjugate $\tilde{q}^+$ can be joined into a doublet satisfying a pseudo-reality condition \cite{Book}, the ``matrix'' part
of the $R$-symmetry generator $\mathcal{R}$ in the superhelicity operator (\ref{superhelicity_analytic}) should vanish on $q^+$ and $\tilde{q}^+$. So we have
\begin{equation}
    \mathbb{W}_{\alpha\dot{\alpha}} q^+ = 0, \ \rightarrow \lambda^{H} = 0. \label{SHq}
\end{equation}
The same is true of course for the ``super $U(1)$-helicity'' defined by eq. \eqref{Rsymmomega}, $\mathbb{W}^{U(1)}_{\alpha\dot{\alpha}}q^+ = 0 \,\rightarrow\, \lambda^{U(1)}_q = 0$.

To determine the helicities of two independent spinorial components of $q^+$, {\it viz.} $D^-_{\hat\alpha}q^+$, we apply  $\mathbb{W}_{\alpha\dot{\alpha}}$ to $ D^{-}_{\hat{\alpha}}q^+$
and use the commutation relations \eqref{DDW}:
\bea
&&    \mathbb{W}_{\gamma\dot{\gamma}} D^{-}_{{\alpha}}q^{+} = (D^{-}_{{\alpha}}\mathbb{W}_{\gamma\dot{\gamma}}
    + [\mathbb{W}_{\gamma\dot{\gamma}}, D^{-}_{{\alpha}}]) q^{+} = -{i}\partial_{\gamma\dot{\gamma}} D^{-}_{{\alpha}}q^{+}\,, \lb{qSpir1} \\
&& \mathbb{W}_{\gamma\dot{\gamma}} \bar{D}^{-}_{{\dot\alpha}}q^{+} = (\bar{D}^{-}_{{\dot\alpha}}\mathbb{W}_{\gamma\dot{\gamma}}
    + [\mathbb{W}_{\gamma\dot{\gamma}}, \bar{D}^{-}_{{\dot\alpha}}]) q^{+} = {i}\partial_{\gamma\dot{\gamma}} \bar{D}^{-}_{{\dot\alpha}}q^{+}\,.  \lb{qSpir2}
\eea
So the helicities of these two spinorial components with respect to the correctly normalized superhelicity operator $\frac12\mathbb{W}_{\gamma\dot{\gamma}}$
are just $\lambda = \frac12$ and $\lambda = -\frac12$, as expected.

In total, we have the following helicity content of hypermultiplet:
\begin{equation}
    q^{+}: \qquad \ket{\lambda=-1/2}, \;\;\ket{\lambda=0}^2, \;\;\ket{\lambda=1/2}, \lb{qCont1}
\end{equation}
{\it i.e.}, two complex scalar fields with zero helicity and two complex Weyl fermionic fields with helicities $\pm 1/2$.
This is precisely content of the on-shell $\mathcal{N}=2$ hypermultiplet (see, e.g., \cite{Book} and refs. therein). All other non-zero projections of $q^+$
are expressed as $x$-derivatives of the basic ones or vanish by the equations of motion \eqref{qplus_equations1} -\eqref{Boxq}. For instance, $q^+|_{\theta =0}$ is $SU(2)_A$
doublet in virtue of the equation $\partial^{++}q^+|_{\theta =0} = 0 \,\rightarrow \,q^+|_{\theta =0} = q^i u^+_i$,  $(D^-_\alpha\bar D^-_{\dot\alpha}q^+)|_{\theta =0} =
-4i \partial_{\alpha\dot\alpha} \partial^{--}q^+|_{\theta =0} = -4i\partial_{\alpha\dot\alpha} q^i u^-_i$ in virtue of eq. \eqref{qplus_equations3}, etc. To summarize, we have the following
independent on-shell superfield projections for the hypermultiplet
\begin{equation}
    q^{+}\,, \quad D^{--}q^+\,, \quad D^-_\alpha q^+, \quad \bar{D}^-_{\dot\alpha}q^+ \,.\lb{qCont2}
\end{equation}
All next projections are either vanishing or expressed through these basic ones. This can be proved using  {\it Lemma} \eqref{Lemma1}. Since $[D^{--}, W_{\alpha\dot\alpha}]q^+ = 0$,
$D^{--}q^+$ possesses the same helicity as $q^+$, {\it i.e.} zero. Both $q^+$ and $D^{--}q^+$, being expanded over harmonics, amount to $SU(2)$ doublet.  The spinor projections
$D^-_{\hat\alpha}q^+$ are ``killed'' by both $D^{\pm\pm}$ and so are $SU(2)_A$ singlets.

These observations can be made more systematic and covariant by evaluating the eigenvalues of the operator $\tilde{C}_I$ \eqref{isospin_operator} for these superfield projectors, without resorting
to their $\theta=0$ reduction. Using the commutation relations \eqref{HarSpin} and the basic equations of motion \eqref{qEqs}, one finds
\bea
\tilde{C}_{I}\,q^+ = \frac{3}{4}\, q^+, \; \tilde{C}_{I}\,(D^{--}q^+) = \frac{3}{4}\, D^{--}q^+, \; \tilde{C}_{I}\,(D^-_{\hat\alpha}q^+) = 0\,,
\; \tilde{C}_{I}\, (D^-_{\alpha}\bar{D}^-_{\alpha}q^+) = \frac{3}{4}\,D^-_{\alpha}\bar{D}^-_{\alpha}q^+,
\eea
where, in deriving the last relation, we have made use of eq. \eqref{qplus_equations3}. So we correctly reproduced the isospins of physical
components (taking into account that  $D^-_{\alpha}\bar{D}^-_{\alpha}q^+ = \frac{i}{4}\partial_{\alpha\dot\alpha}\,D^{--}q^+$). Denoting physical states as
$\ket{\lambda, I, g_0, {\cal I}}$ where $I$ is the isospin, we have the following on-shell physical content of $q^+$:
\bea
 q^{+}: \qquad \ket{1/2, 0, 0, 0},\quad \ket{0, 1/2, 0, 0}^2,\quad \ket{-1/2, 0, 0, 0}\,. \lb{qContent}
\eea

\subsection{$\omega$  hypermultiplet}

The analytic real superfield $\omega$ possesses the following two equivalent forms of the equation of motion:

\begin{equation}\label{omegamotion}
 (a)\;\;   (D^{++})^2 \omega = 0,  \quad \leftrightarrow \quad (b)\;\;(D^{--})^2 \omega = 0\,.
\end{equation}
The equation $(a)$ can be obtained from the off-shell action
\bea
S_\omega \propto \int d\zeta^{(-4)}\, D^{++}\omega D^{++}\omega\,.
\eea

In the same way as for $q^{+}$, it can be shown that eqs. \eqref{omegamotion}
imply the following descendants
\bea
&&    D^{--}  {D}^{-}_{\hat{\alpha}} \omega = 0\,, \label{omega_1st}\\
 &&   (D^-)^2 \omega = 0, \quad (\bar{D}^-)^2 \omega = 0, \label{omega_2}\\
 &&   \Big(\partial_{\alpha\dot{\alpha}}D^{--} -\frac{i}{4}  D^{-}_{\alpha}\bar{D}^{-}_{\dot{\alpha}}\Big) \omega = 0, \label{omega_3} \\
 &&   \partial_{\alpha\dot{\alpha}}{D}^{-{\alpha}} \omega = \partial_{\alpha\dot{\alpha}}\bar{D}^{-\dot{\alpha}} \omega = 0, \label{omega_4}\\
&&    \Box \omega = 0\,. \label{BoxOmega}
\eea

Eqs. \eqref{omega_4}, \eqref{omega_2} together with those trivially following from the Grassmann analyticity conditions,
$$
\partial_{\alpha\dot\alpha}{D}^{+{\alpha}} \omega = \partial_{\alpha\dot{\alpha}}\bar{D}^{+\dot{\alpha}} \omega = 0, \quad(D^+)^2 \omega = 0, \quad
(\bar{D}^+)^2 \omega = 0\,, (D^+D^-)\omega = (\bar{D}^+\bar{D}^-)^2 \omega = 0\,,
$$
are just avatars of the generic admissible on-shell constraints \eqref{Casimir/Generators/Massless/Qconstraints}, \eqref{Casimir/Generators/Massless/Qconstraints2}.

Before discussing implications of the super-isohelicity operators $\mathbb{W}^{\pm\pm}, \mathbb{W}^{0}$ to the given system, let us firstly
find out the helicity contents of $\omega$.

\textbf{Helicity contents.}
Starting from $\omega$, one can construct the following set of independent superfield projections:
\begin{equation}\label{omega_components}
    \omega, \;\;\; D^{\pm\pm}\omega, \;\;\;  D^{++}D^{--}\omega, \;\; \; D^{-}_{\hat{\alpha}}\omega, \;\;\;  D^{++}D^{-}_{\hat{\alpha}}\omega = D^{-}_{\hat{\alpha}}D^{++}\omega.
\end{equation}
Using the dynamical equations it is possible to check that this set indeed exhausts all possibilities: any other possible non-vanishing projection is either harmonic or $x$-derivatives
of entries in \eqref{omega_components}, or just zero. For instance, acting by $D^{++}$ on eq. \eqref{omega_3}, we obtain
$$
D^{-}_{{\alpha}}\bar{D}^{-}_{\dot{\alpha}}D^{++}\omega = 4i \partial_{\alpha\dot\alpha} \big(\omega - D^{++}D^{--}\omega\big)\,.
$$

Since the analytic superfield $\omega$ is real, the $R$-symmetry operator \eqref{Rsymmomega} is vanishing on $\omega$. Also vanishing is the superhelicity operator (\ref{superhelicity_analytic})
\bea
\mathbb{W}_{\alpha\dot{\alpha}}\omega = \mathbb{W}^{U(1)}_{\alpha\dot{\alpha}}\omega = 0\: \Rightarrow \; \lambda^H = \lambda^{U(1)} = 0\,.
\eea

To determine the helicity content of $\omega$, we can proceed like in the $q^+$ case.
We hit $\omega$ with spinor derivatives,
\begin{equation}
    \omega, \quad D^{-}_{\hat{\alpha}} \omega, \quad  D^{-}_{\alpha}\bar{D}^{-}_{\dot{\alpha}}\omega,\lb{Projomega}
\end{equation}
and act on these projections by the operator $\mathbb{W}_{\alpha\dot\alpha}$. For spinor projections we find the helicities $\lambda = \pm 1/2$ and for
$\omega$ itself and its vector projection, $\lambda =0$.
This is quite natural because, in virtue of eq. \eqref{omega_3}, the latter projection is a gradient of the scalar component $D^{--}\omega|_{\theta=0}$ with zero helicity.
So the helicity states are:
\begin{equation}
    \ket{\lambda=-1/2}, \quad \ket{\lambda=0}^2, \quad \ket{\lambda=1/2}\,.
\end{equation}
And then we need to hit this set by all possible combinations of $D^{\pm\pm}$ operators and collect a linearly independent set. Note that there are four independent projections with zero helicity:
\be
\omega, \quad D^{\pm\pm}\omega, \quad D^{++}D^{--}\omega\,,
\ee
which reflects the property that the scalar fields of $\omega$ are grouped just into the singlet and triplet of $SU(2)_A$. The spinorial components are grouped into $SU(2)_A$ doublets,
\bea
D^-_{\hat\alpha}\omega\,, \quad D^-_{\hat\alpha}D^{++}\omega\,,
\eea
as distinct from $q^+$, the spinor projections of which are $SU(2)_A$ singlets. All these statement can be easily   reproduced by applying the isospin-square operator $\tilde{C}_I$
as in the case of $q^+$ hypermultiplet. We have
\bea
&& \tilde{C}_I\, (D^{\pm\pm}\omega,\; D^{++}D^{--}\omega)  =  2 \,(D^{\pm\pm}\omega, \; D^{++}D^{--}\omega)\,, \quad \tilde{C}_I\,\omega_0 = 0\,, \\
&&\tilde{C}_I\,(D^-_{\hat\alpha}\omega) = \frac{3}{4}\,D^-_{\hat\alpha}\omega \,, \quad \tilde{C}_I\,(D^-_{\hat\alpha}D^{++}\omega) = \frac{3}{4}\,D^-_{\hat\alpha}D^{++}\omega\,, \lb{ContOm}
\eea
where
\bea
\omega_0 :=\omega - \frac12 D^{++}D^{--}\omega\,, \quad D^{\pm\pm}\omega_0 = 0\,. \nonumber
\eea

\textbf{Super-isohelicity.} Let us see what kind of representation of the algebra \eqref{Wops_commutation} we encounter in the present case. Once again, we can define
the ``vacuum state'' by the condition
\bea
\mathbb{W}^{--}_{\alpha\dot\alpha} \omega = 0\,, \lb{W--Om}
\eea
which is just the equation of motion \eqref{omega_3}. Also, it is direct to check that
\bea
&& \mathbb{W}^{++}_{\alpha\dot\alpha} \omega := \omega^{++}_{\alpha\dot\alpha} = D^{++}\partial_{\alpha\dot\alpha} \omega\,, \quad   \mathbb{W}^{0}_{\alpha\dot\alpha} \omega =
- \partial_{\alpha\dot\alpha} \omega\,,  \lb{Aom} \\
&& \mathbb{W}^{++}_{\alpha\dot\alpha} \mathbb{W}^{++}_{\beta\dot\beta}\omega \sim (D^{++})^2 \omega = 0\,. \lb{Bom}
\eea
We also have
\bea
\mathbb{W}^{0}_{\alpha\dot\alpha} \omega^{++}_{\beta\dot\beta} = \partial_{\alpha\dot\alpha}\omega^{++}_{\beta\dot\beta}\,, \quad \mathbb{W}^{--}_{\alpha\dot\alpha} \omega^{++}_{\beta\dot\beta} =
- \partial_{\alpha\dot\alpha} \partial_{\beta\dot\beta}\omega\,,  \lb{Aom}
\eea
where we used the commutation relations \eqref{Wops_commutation} and the fact that the general on-shell conditions
\eqref{Casimir/Generators/Massless/Qconstraints}, \eqref{Casimir/Generators/Massless/Qconstraints2} are satisfied in the present case like in the previous $q^+$ case.
So we obtained that the ``module'' $(\omega, \,\omega^{++}_{\alpha\dot\alpha})$ is closed under the action of generators
$\mathbb{W}^{0}_{\alpha\dot\alpha}, \mathbb{W}^{\pm\pm}_{\alpha\dot\alpha}$.

For the numbers $\mathcal{I}$ and $g_0$ we obtain:
\begin{align}
    \mathbb{W}^{0}_{\alpha\dot{\alpha}}\omega = - \partial_{\alpha\dot{\alpha}}\omega\,, \quad \rightarrow \quad g_0 =-1\,,\lb{Omg0}
\end{align}
and
\begin{equation}
    C^{[I]}_{\alpha\beta\dot{\alpha}\dot{\beta}} \omega \simeq \frac{3}{4} \partial_{\alpha{\dot{\alpha}}}\partial_{\beta\dot{\beta}} \omega \quad \rightarrow \quad {\cal I} = \frac12\,.
\end{equation}
Defining the states as in the $q^+$ case, $\ket{\lambda, I, g_0, {\cal I}}$, we can now present the full set of states of the on-shell $\omega$ supermultiplet as :
\begin{align}
\omega: \quad \ket{\lambda, I, g_0, {\cal I}} =& \ket{0, \,0, \, -1, \, 1/2}, \quad \ket{0, 1, -1, 1/2} \;\; ({\bf {\rm bosonic }}), \\
& \ket{1/2, \, 1/2, \, -1,  1/2}, \quad \ket{-1/2, \, 1/2, -1, 1/2} \; \;({\bf {\rm fermionic}}). \lb{OmContent}
\end{align}

Like in the $q^+$  case, the numbers $g_0$ and  ${\cal I}$  characterize the $\omega$ multiplet as the whole, while $\lambda$ and $I$ are different for its various superfield components.
It is instructive to explicitly give the relevant set of the component fields:
\begin{equation}
\omega: \quad \omega(x), \;\; \omega^{(ij)}(x), \;\;\psi^i_\alpha(x),\;\; \bar{\psi}^i_{\dot{\alpha}}(x)\,.
\end{equation}

\subsection{${\cal N}=2$ super Maxwell theory}

The analytic gauge prepotential $V^{++}$ in the abelian case has the following action and equation of motion \cite{Book}:
\bea
&& S_V \sim \int d^4x d^8\theta du \, V^{++} V^{--}, \label{vppAction} \\
 &&    (D^+)^4 V^{--} = 0,\label{vppmotion}
\eea
where the non-analytic gauge connection $V^{--}$ is related to $V^{++}$ by the harmonic flatness condition:
\begin{equation}
    D^{++}V^{--} = D^{--}V^{++}. \label{FlatVV}
\end{equation}
In terms of $V^{\pm\pm}$ the underlying gauge transformations with the analytic gauge parameter $\Lambda$ are realized as
\bea
\delta V^{\pm\pm} = D^{\pm\pm}\Lambda\,.\lb{GaugeVtr}
\eea

One way to deal with this system is to fix  the gauge \cite{SupergravityMotionEquations}:
\begin{equation}\label{vppgauge}
    D^{++}V^{++} = 0\; \rightarrow \; V^{--} = \frac{1}{2}(D^{--})^2 V^{++}\,.
\end{equation}
It is equivalent to the condition
\bea
(D^{--})^3 V^{++} = 0\,, \lb{Gauge--}
\eea
which can be proved using {\it Lemma} \eqref{Lemma1}. Employing these equivalent forms of the gauge condition together with the
dynamical equation \eqref{vppmotion}, one can derive their various corollaries, in particular,
\begin{equation}\label{vppmass}
    \Box V^{++} = 0\,,
\end{equation}
which means that the off-shell Casimir operators defined in section 3 become singular on shell in the case under consideration,
and so we need another type of invariant operators to treat the massless case, like in the cases of the free $q^+$ and $\omega$ hypermultiplets.

Like in the non-supersymmetric case, in order to define the physically important quantities with fixed super-helicity and super-isohelicities,
it is preferable to deal with the gauge-invariant quantities, the covariant superfield strengths which in the harmonic prepotential formulation
are defined as:
\begin{equation}\label{Casimir/ActionSuperfield/Wexpression}
W=-\frac{1}{4}(\bar{D}^+)^2V^{--}\,, \qquad \overline{W} = -\frac{1}{4}(D^+)^2 V^{--}\,.
\end{equation}
The covariant strengths obey the following off-shell constraints
\bea
&& D^{\pm\pm}W = D^{\pm\pm}\bar W = 0\,, \quad \bar{D}^+_{\dot\alpha} W = \bar{D}^-_{\dot\alpha} W =0\,, \;  {D}^+_{\alpha} \bar W = {D}^-_{\alpha} \bar W =0\,, \label{Off1}\\
&&(D^+)^2 W = (\bar{D}^+)^2\bar{W}, \label{Off2}
\eea
which follow from their definition, analyticity of $V^{++}$ and the flatness condition \eqref{FlatVV}.

To show that the case of ${\cal N}=2$ gauge theory can be treated quite analogously to the free cases considered before, it is convenient to directly deal with
the covaraiant strengths $W, \bar W$ defined by the off-shell constraints \eqref{Off1} and \eqref{Off2}. The latter
is just ${\cal N}=2$ generalization of the standard Bianchi identity for the Maxwell strength ${\cal F}^{(\alpha\beta)}, \bar{\cal F}^{(\dot\alpha\dot\beta)}$ \footnote{Actually,
representing through harmonic potentials is imperative when considering interactions with matter superfields, while at the free-field level we can proceed from the definitions
\eqref{Off1} and \eqref{Off2}, with the free action as integrals over chiral ${\cal N}=2$ superspaces, $S_{free} \sim \int d^4 x_L d^4\theta \, W^2 + c.c.$\,.}.

Since the operators $(D^{+})^2$ and $(\bar D^+)^2$ possess non-zero $R$ symmetry weights (under ``passive'' realization of $R$-symmetry as transformations of Grassmann coordinates),
for $R$-covariance of \eqref{Off2} we are led to ascribe non-zero $R$-weights to $W$ and $\bar W$:
\be
{\cal R} W = -2i W\,, \quad  {\cal R} \bar{W} = 2i \bar{W}\,.\lb{Weights}
\ee
These weights can be directly derived from the representation \eqref{Casimir/ActionSuperfield/Wexpression} by noting that $RV^{--} = R^{\theta}V^{--}$
(see \eqref{SplR} and \eqref{Rgen}
for definitions) and then commuting $R^{\theta}$ with $(D^{+})^2$ and $(\bar{D}^+)^2$.

To avoid a possible misunderstanding, we note that in the ``active'' realization of $R$-symmetry transformations, when spinor coordinates and derivatives do not transform at all,
these weights appear due to the non-commutativity between the differential part of the full $R$-symmetry generator acting on $V^{--}$ and the operators $(D^{+})^2$ and $(\bar{D}^+)^2$.
The prepotentials $V^{\pm\pm}$ themselves carry no any weight.

The equation of motion of the abelian ${\cal N}=2$ gauge theory can be written in terms of $W, \bar{W}$ as
\bea
(D^+)^2 W =(\bar{D}^+)^2\bar{W} = 0\,, \quad (D^+ D^-)W =(\bar{D}^+ \bar{D}^-)\bar{W} = 0\,,\label{OnW}
\eea
while other constraints in \eqref{Off1}, \eqref{Off2} remain unaffected. Eqs. \eqref{OnW} combine dynamical equations with Bianchi identities and so
are genuine ${\cal N}=$ analogs of the Bargmann-Wigner equations $\partial_{\alpha\dot\alpha}{\cal F}^{(\alpha\beta)}
= \partial_{\alpha\dot\alpha}{\cal F}^{(\dot\alpha\dot\beta)} = 0\,.$ Now, acting on the first equation in \eqref{OnW} by $\bar{D}^-_{\dot\alpha}$
and using the chirality of $W$, we obtain the relevant Dirac equations
\bea
\partial_{\beta\dot\alpha} D^{+\beta} W = 0\,, \quad \partial_{\beta\dot\alpha}\bar{D}^{+\beta} \bar{W} = 0\,.  \lb{WDir}
\eea
Applying $\bar{D}^-_{\dot\rho}$ to the first of these equations and using the chirality of $W$ (and anti-chirality of $\bar W$) once more, we obtain
\bea
\Box W  = 0, \;\quad \Box \bar{W}  = 0. \lb{Wbox}
\eea
Acting on \eqref{WDir} by $D^{--}$ and using $D^{--}W =0$ we also  have
\bea
\partial_{\beta\dot\alpha} D^{-\beta} W = \partial_{\beta\dot\alpha}\bar{D}^{-\dot\alpha} \bar{W} = 0\,. \lb{WDir2}
\eea
Dynamical equations \eqref{WDir2}, \eqref{OnW} combined with their off-shell counterparts implied by the chirality conditions,
\bea
\partial_{\beta\dot\alpha} \bar{D}^{+\dot\alpha} W = 0 \,,  \partial_{\beta\dot\alpha}{D}^{+\beta} \bar{W} = 0, \;
(\bar{D}^+)^2 W =({D}^+)^2\bar{W} = (\bar{D}^+\bar{D}^-)W =({D}^+ D^-)\bar{W} = 0\,, \lb{WDirTriv}
\eea
amount to the generic on-shell constraints \eqref{Casimir/Generators/Massless/Qconstraints}, \eqref{Casimir/Generators/Massless/Qconstraints2}.\\

\textbf{Super-isohelicities.}
Having the full set of the Dirac-type on shell constraints on $W, \bar W$, we can calculate the relevant values of $\mathcal{I}$ and $g_0$.
We obtain:
\begin{align}
    \mathbb{W}^{0}_{\alpha\dot{\alpha}} W = 0 \quad  \Rightarrow \quad g_0 = 0.
\end{align}
It is also easy to show that
\begin{equation}
    \mathbb{W}^{\pm\pm}_{\alpha\dot{\alpha}} W = 0 \ \Rightarrow \ \mathcal{I} = 0.
\end{equation}

\textbf{Superhelicity and multiplet contents.}
It is direct to find
\be
\mathbb{W}_{\alpha\dot\alpha} (W, \bar W) = 0\,, \quad \Rightarrow \quad \lambda^H_{(W)} = 0\,.
\ee

The full set of independent superfield projections is obtained through acting on $W$ by $D^{+}_{\alpha}$ and $D^{-}_{\alpha}$,
\begin{equation}
    W, \quad D^{\pm}_{\alpha}W, \quad D^{-}_{(\alpha}D^{+}_{\beta)}W\,.\lb{setW}
\end{equation}

Using the definition \eqref{superhelicity_analytic} and \eqref{Weights} we obtain
\begin{eqnarray}
&&    \mathbb{W}_{\alpha\dot{\alpha}} W = -i \partial_{\alpha\dot{\alpha}}W  + i\partial_{\alpha\dot\alpha} W = 0 \;\rightarrow \; \lambda^{H} = 0, \nonumber \\
&& \mathbb{W}_{\alpha\dot{\alpha}}\,D^{\pm}_\beta W = -i D^{\pm}_\alpha \partial_{\beta\dot\alpha}  W  \simeq -i D^{\pm}_\beta \partial_{\alpha\beta} W\,, \nonumber \\
&& \mathbb{W}_{\alpha\dot{\alpha}}\,\bar{D}^{\pm}_{\dot\beta} \bar{W} = i \bar{D}^{\pm}_{\dot\alpha} \partial_{\alpha\dot\beta}  \bar{W}
\simeq i \bar{D}^{\pm}_{\dot\beta} \partial_{\alpha\dot\alpha}  \bar{W}\,, \nonumber \\
&& \mathbb{W}_{\alpha\dot{\alpha}} D^+_{(\gamma} D^-_{\beta)} W \simeq -2i D^+_{(\gamma} D^-_{\beta)} \partial_{\alpha\dot\alpha} W \; \;({\rm and \; c.c.}).
\end{eqnarray}

So $W, \bar W$ have zero helicity, $D^{\pm}_\alpha W$ and $\bar{D}^{\pm}_{\dot\alpha} \bar W$ have helicities $1/2$ and $-1/2$ while $D^+_{(\gamma} D^-_{\beta)} W$
and its conjugate carry helicities $\pm 1$. Using the isospin-square operator $\tilde{C}_I$ it is also easy to check that the scalar and vector superfields in the set
\eqref{setW} carry zero $I$, while the spinor superfields carry isospin $I=1/2$. This is just the content of the on-shell ${\cal N}= 2$ gauge multiplet.

In terms of the on-shell sates $\ket{\lambda,  I, g_0, \tilde{I}}$ we obtain the following set of independent states:
\begin{equation}
    W: \quad \big( \ket{-1, 0,0,0}, \quad \ket{-1/2, 1/2, 0, 0}, \quad \ket{0, 0, 0, 0}^2, \ket{1/2, 1/2, 0,0}, \ket{1, 0, 0, 0} \big).
\end{equation}

The super $U(1)$ helicity defined by eqs. \eqref{Rsymmomega}, \eqref{U1hel}  plays some role in the present case: it discriminates between two $\ket{0, 0, 0}^2$ states,
so that $\lambda^{U(1)}_W =1$ and  $\lambda^{U(1)}_{\bar W} =-1$. Also, the non-zero superfield projections $D^\pm_\alpha W$ and $D^+_{(\alpha}D^-_{\beta)}W$ have non-zero $U(1)$
helicities +2 and +1, respectively, while their conjugates -2 and -1.

\section{Summary of Casimir operators and outlook}

\noindent{\textbf{Casimir operators in massive off-shell case}}

\begin{itemize}
    \item Square of the momentum:
    \begin{equation}
    P^n P_n = -\partial^n\partial_n = -\Box; \nonumber
    \end{equation}
    \item Superspin operator
     \begin{align}
    C_S &= 2\Box(\hat{W}^{\dot{\alpha}\alpha}\hat{W}_{\dot{\alpha}\alpha}), \nonumber \\
 \hat{W}_{\dot{\alpha}\alpha} &= W_{\dot{\alpha}\alpha} + \frac{1}{8}[Q^i_{\alpha},\bar{Q}_{i\dot{\alpha}}] - \frac{1}{4\Box}\left[Q^i_{\gamma},\bar{Q}_{i\dot{\gamma}}\right]
    \partial^{\dot{\gamma}\gamma}\partial_{\alpha\dot{\alpha}} \nonumber \\
    &= \mathcal{W}_{\alpha\dot{\alpha}} +\frac{1}{4}(D^-_{\alpha}\bar{D}^{+}_{\dot{\alpha}} + \bar{D}^-_{\dot{\alpha}}D^+_\alpha)
    - \frac{1}{2\Box}(D^-_{\gamma}\bar{D}^{+}_{\dot{\gamma}} + \bar{D}^-_{\dot{\gamma}}D^+_\gamma)\partial^{\dot{\gamma}\gamma}\partial_{\alpha\dot{\alpha}}\,,\nonumber
    \end{align}
    where $\mathcal{W}_{\alpha\dot{\alpha}}$ is the matrix part of $W_{\alpha\dot\alpha}$;

    \item Superisospin operator:
    \begin{equation}
    \begin{split}
    C_I &= \hat{T}^i_j \hat{T}^j_i, \nonumber\\
    \hat{T}^i_j &= T^i_j - \frac{1}{2\Box}[Q^{i\alpha},\bar{Q}^{\dot\alpha}_j]\partial_{\alpha\dot\alpha}
    + \frac{1}{4\Box}\delta^i_j[Q^{k\alpha},\bar{Q}^{\dot\alpha}_k]\partial_{\alpha\dot\alpha}\,,  \nonumber\\
    \end{split}
    \end{equation}
    or, in terms of covariant derivatives,
    \begin{equation}
    C_I = \frac{1}{4}\left(\nabla^0\right)^2 + \frac{1}{2}\left\{\nabla^{++},\nabla^{--}\right\}, \nonumber
    \end{equation}
    where
    \begin{eqnarray}
     &&   \nabla^{++} = D^{++} + \frac{i}{\Box}(D^{\alpha+}  \bar{D}^{+\dot\alpha})\partial_{\alpha\dot\alpha}, \nonumber \\
     &&   \nabla^{--} = D^{--} - \frac{i}{\Box}(D^{\alpha-}  \bar{D}^{-\dot\alpha})\partial_{\alpha\dot\alpha}, \nonumber \\
     &&   \nabla^{0} = (D^0 - 2) - \frac{i}{\Box}(D^{-}_{\alpha}\bar{D}^{+}_{\dot\alpha} - \bar{D}^-_{\dot\alpha}  D^+_\alpha)\partial^{\dot{\alpha}\alpha};\nonumber
      \end{eqnarray}

   \item Super-U(1)charge operator:
    \begin{align}
    C_{U(1)} &= -i\hat{R} = -iR + \frac{i}{2\Box}[Q^{i\alpha},\bar{Q}^{\dot\alpha}_i]\partial_{\alpha\dot\alpha} \nonumber\\
    &=-i\mathcal{R} -\frac{i}{\Box}(\bar{D}^-_{\dot\alpha}D^+_{\alpha} + D^{-}_{\alpha}\bar{D}^{+}_{\dot\alpha})\partial^{\dot\alpha\alpha}.\nonumber
    \end{align}
\end{itemize}

\begin{eqnarray}
&& C_S\Phi^{(q)} = \Box^2 S(S+1) \Phi^{(q)}, \quad C_I \Phi^{(q)}_{(I)} = I(I+1) \Phi^{(q)}_{(I)}, \quad C_{U(1)} \Phi^{(q)} = R \Phi^{(q)}\,, \nonumber \\
&& I = \biggr|\frac{q-2}{2}\biggr|, \quad \nabla^{++}\Phi^{(q)}_{(I)} = 0 \;\;{\rm  for} \;\;q \geq 2\,, \;\;\nabla^{--}\Phi^{(q)}_{(I)} = 0 \;\;{\rm  for} \;\;q =0, q=1 \,.
\end{eqnarray}

In the Table 1 we collect values of Casimir operators for off-shell superfields:
\begin{table}[h!]
    \begin{tabular}{|l|lll|}
        \hline
        & ${\cal R}$ & $S$ & $I$                             \\ \hline
        $q^+$    & 0   & 0   & $\frac{1}{2}, \frac{3}{2}, ...$ \\
        $\omega$ & 0   & 0   & $1, 2, ...$                     \\
        $V^{++}$ & 0   & 0   & $\mathbf{0}, 1, 2, ...$         \\ \hline
    \end{tabular}
    \centering
    \caption{Values of Casimir operators on off-shell superfields}
\end{table}

\noindent{\textbf{Massless case} }

\begin{itemize}
    \item Superhelicity:
      \begin{align}
    \mathbb{W}_{\alpha\dot{\alpha}}=&
    \mathcal{W}_{\alpha\dot{\alpha}} +\frac{1}{4}(D^-_{\alpha}\bar{D}^{+}_{\dot{\alpha}}
    + \bar{D}^-_{\dot{\alpha}}D^+_\alpha) - \frac12{\cal R}\partial_{\alpha\dot{\alpha}} \simeq \lambda^H P_{\alpha\dot{\alpha}}; \nonumber
    \end{align}

\item Isospin-square operator:
 \begin{align}
  \tilde{C}_I = \frac{1}{4}\,{(D^0)^2} + \frac{1}{2}\,\{D^{++}, D^{--}\};
\end{align}

    \item Super $U(1)$ helicity:
    \begin{align}
    \mathbb{W}^{U(1)}_{\alpha\dot{\alpha}} &= \mathcal{R}\partial_{\alpha\dot{\alpha}}
    - \frac{1}{4}(D^-_{\alpha}\bar{D}^{+}_{\dot{\alpha}} + \bar{D}^-_{\dot{\alpha}}D^+_\alpha) \simeq \lambda^{U(1)} P_{\alpha\dot{\alpha}}; \nonumber
    \end{align}
    \item Super-isohelicities:
     \begin{eqnarray}
    &&\mathbb{W}^{\pm\pm}_{\alpha\dot{\alpha}} \simeq D^{\pm\pm}\partial_{\alpha\dot{\alpha}} \pm \frac{i}{4}D^\pm_\alpha, \bar{D}^\pm_{\dot{\alpha}}\,, \nonumber\\
    && \mathbb{W}^{0}_{\alpha\dot{\alpha}} \simeq D^0 \partial_{\alpha\dot{\alpha}} -  \frac{i}{4}(D^+_\alpha \bar{D}^-_{\dot{\alpha}} + D^-_\alpha \bar{D}^+_{\dot{\alpha}}) \nonumber\\
    &&= (D^0 - 1)\partial_{\alpha\dot{\alpha}} - \frac{i}{2}(D^-_\alpha \bar{D}^+_{\dot{\alpha}} - \bar{D}^-_{\dot{\alpha}} D^+_\alpha) \simeq g_0\partial_{\alpha{\dot{\alpha}}}. \nonumber
    \end{eqnarray}
\end{itemize}
\newpage

In the Table 2 we present the values of superhelicity $\lambda^{H}$, isospin $I$ and super-isohelicity $(g_0, \mathcal{I})$,
together with the relevant massless multiplet contents:
\begin{table}[h!]
    \begin{tabular}{|l|lll|}
        \hline
        & $\lambda^{H}$ & $(g_0, \mathcal{I})$ & $\ket{\lambda, {I}}$                                    \\ \hline
        $q^+$                                                              & 0              & $(0, 0)$             & $\ket{0,1}, \;\;\ket{\pm 1/2, 0}$                 \\
        $\omega$                                                           & 0              & $(-1, 1/2)$        & $\ket{0,0}, \;\;\ket{0,1}, \;\; \ket{\pm1/2, 1/2}$ \\
        $W, \bar{W}$                                           & 0         & $(0, 0)$             & $\ket{0,0}^2, \;\;\ket{\pm1/2,1/2}, \;\;\ket{\pm1, 0}$           \\\hline
    \end{tabular}
    \centering
    \caption{On-shell superfields}
\end{table}
\vspace{0.5cm}

Among problems for further study we mention generalization of the present construction to the linearized $4D, {\cal N}=2$ supergravity (superspin 2) and $4D, {\cal N}=2$
higher  spin theories \cite{BIZ1,SupergravityMotionEquations} in the harmonic approach. It is expected that in the massless case the results of ref. \cite{SupergravityInvariants}
will be of great help. Also it would be interesting to extend the methods developed here for the flat ${\cal N}=2$ superspace to its AdS$_4$ analogue \cite{AdS,AdS2}.

One more task is to apply these techniques to the central charge extension of ${\cal N}=2$ supersymmetry (recall subsection 2.5). This case is of essential interest since, e.g., hypermultiplets
within such a framework can acquire a mass through the famous Scherk-Schwarz mechanism. To the best of our knowledge,  nobody studied the role of superspin, superisospin and other Casimir operators
in the context of this central-charge extended ${\cal N}=2$ supersymmmetry. The relevant harmonic superspace formalism should presumably resemble the $5D$ one of ref. \cite{5D}, or its full-fledged $6D$ version.
It is curious that
the enlarged harmonic derivatives $\nabla^{\pm\pm}, \nabla^0$ admit an immediate  generalization to the central-charge case through the proper insertion of the operators
$(D^\pm)^2\frac{\partial}{\partial x^5}, \;\;(D^+D^-)\frac{\partial}{\partial x^5}, \;\;(\bar{D}^{\pm})^2\frac{\partial}{\partial x^5},\;\;(\bar{D}^{+}\bar{D}^{-})\frac{\partial}{\partial x^5}$
in the numerators of the definitions\eqref{Casimir/CovDerExpr/Nablas} and
the change $\Box \rightarrow \Box + b (\frac{\partial}{\partial x^5})^2$, with $b$ being some number, in the denominators.

\acknowledgments


The present study  was supported by the Foundation for the Advancement of Theoretical Physics and
Mathematics ``BASIS'',  grant \verb|#| 25-1-1-10-4. We thank Ioseph Buchbinder and Nikita Zaigraev
for interest in the work and useful comments. E.I. is indebted to Mikhail Vasiliev for an important remark.

 \appendix

\section{$SU(2)$ multiplets in the harmonic formalism}

Here we deal with the pure harmonic $S^2\sim SU(2)_A/U(1)_A$ stuff and use the harmonic derivatives
\begin{align}
    \partial^{\pm\pm} = u^{\pm i} \frac{\partial}{\partial u^{\mp i}}, \qquad \partial^0
    = u^{+i}\frac{\partial}{\partial u^{+i}} - u^{-i}\frac{\partial}{\partial u^{-i}}, \label{HDer}
\end{align}
which satisfy the $su(2)$ algebra relations
\beq
[\partial^{++}, \partial^{--}] = \partial^0\,, \quad [\partial^0,\partial^{\pm\pm}] = \pm 2\partial^{\pm\pm}\,.\label{A1}
\eeq

The functions $f^{+a}(u)$ on the coset $S^2$ are assumed to be classified according to their external harmonic $U(1)_A$ charges $a$ (the number of indices $+$)
\beq
\partial^0 f^{+a} = a\,f^{+a}\,, \quad a\geq 0\,.\label{A5}
\eeq
Eq. \eqref{A5} implies that the harmonic expansion of $f^{+a}(u)$ contains an infinite number of the harmonic monomials, all having the charge $+a$
\beq
f^{+ (a)} = f^{(i_1\ldots i_a)} u^+_{(i_1}\cdots u^{+}_{i_a)} +  f^{(i_1\ldots i_a j_{a+1}j_{a +2})} u^+_{(i_1}\cdots u^{+}_{i_a}u^{+}_{j_{a+1}} u^{-}_{j_{a+1})} + \dots\,. \label{A6}
\eeq
The functions $f^{+(a)}$ are assumed to be square-integrable on $S^2$. The functions with negative $U(1)$ charges can be obtained by acting
of the appropriate degree of $\partial^{--}$ on some function with a non-negative charge, e.g.,
$$
f^{-} = \partial^{--} f^+\,, \quad  f^{-2} = \partial^{--} f^0\,,
$$
so it suffices to deal with the non-negatively charged functions \footnote{Recall footnote 5.}.
 Also, any function with an even charge $a=2n$, or odd one $a= 2n +1$, can be reproduced
    as \beq f^{+ 2n} = (\partial^{++})^n f^0\,, \quad f^{+ (2n +1)} = (\partial^{++})^n f^+\,, \eeq and
so the functions $f^0 \equiv \omega$ and $f^+$ can be regarded as the basic ones. In general, $f^{+ a}$ is complex with
respect to the ``tilde-conjugation'' (the product of the ordinary complex conjugation and Weyl
reflection on $S^2$) which is squared to $-1$ on the harmonic variables. For even $a$ one can impose the reality condition
\beq
\widetilde{f^{+ 2n}} = f^{+ 2n}\,,\lb{Realf}
\eeq
which reduces the number of independent components in
$f^{+ 2n}$ (integer isospins) by twice. In what follows, we always assume \eqref{Realf} for $a=2n$. For $a= 2n+1$ such a reduction is impossible, so the components of $f^{+ (2n+1)}$ (half-integer isospins) are intrinsically
complex. The harmonic derivatives $\partial^{\pm\pm},
\partial^{0}$ are real with respect to the tilde-conjugation \footnote{These are real with respect to the ordinary conjugation too, but here we will be interested in the tilde conjugation because the basic notions of ${\cal
        N}=2$ HSS approach are covariant just with respect to the latter.}. There will be also useful the following statements (which are proved using the harmonic expansions of $f^{+a}$):
\bea
&&\partial^{--} f^{+ a} = 0
\; \Rightarrow \; f^{+a} = 0 \;\; {\rm for}\;\;  a > 0\,, \nn
&&\partial^{--} f^{0} = 0\;\; ({\rm or, \;equivalently}, \;\partial^{++} f^{0} = 0)\; \; \Rightarrow \; f^{0} = const\,. \label{Lemma}
\eea

\noindent{\bf Digression}.  Before moving on, note that such functions on $S^2$, with an arbitrary $U(1)$ acting
on the index $a$, from the mathematical point of view are called ``sections'' and the monomials constructed out of the isospinor harmonic variables $u^\pm_i$ are sometimes
dubbed the ``weighted harmonics'' \cite{WuYang,Glenn}.
The basic distinction  of the approach which was introduced in \cite{Book} and to which we stick here,
is that we use the parametrization-independent description of the 2-sphere
$S^2$ by ``diades'' $u^\pm_i, \;u^{+i}u^-_i = 1,$ globally defined over the whole sphere $S^2$. The use of the derivatives on $S^2$ in the very simple form \eqref{HDer} also radically simplifies many formulas which
look very bulky in any particular explicit
parametrization of $S^2$.
What is frequently called ``functions on $S^2$'' in our language corresponds to considering only those $f^{+n}$ which have zero charge, $n=0$. Such functions evidently carry only integer-isospin multiplets
with ${\bf s} = 0, 1, 2, \ldots $ and their harmonic expansion actually goes over powers of zero-charge (triplet) harmonic monomials $u^+_{(i}u^-_{k)}$.

The $SU(2)$ Casimir operator in our approach is expressed through harmonic derivatives as
\beq
C= \frac12\, \{\partial^{++}, \partial^{--}\} + \frac14 (\partial^0)^2. \label{CasimSU2}
\eeq
It commutes with both the $SU(2)$ generators realized on the doublet indices of $u^{\pm}_i$ and the derivatives $\partial^{\pm\pm}, \partial^0\,.$
The useful consequences of the definition \eqref{CasimSU2} and the commutation relations \eqref{A1} are
\beq C = \partial^{--}\partial^{++} +\frac12  \partial^0\, \big( 1 + \frac12 \partial^0 \big), \quad
C = \partial^{++}\partial^{--} - \frac12  \partial^0 \,\big( 1 -\frac12 \partial^0 \big). \label{Cas12}
\eeq

Let us constrain $f^{+a}$ as
\beq
\partial^{++} f^{+a} = 0 \quad \Rightarrow \quad f^{+a} := \hat{f}^{+a}_{{\bf s} = a/2}  = f^{(i_1\ldots i_a)} u^+_{(i_1}\cdots u^{+}_{i_a)}\,.
\eeq
It is straightforward to check that
\beq
C\, \hat{f}^{+a}_{{\bf s} = a/2}  = {\bf s} \big( 1 + {\bf s} \big) \hat{f}^{+a}_{{\bf s} = a/2} \,,
\eeq
{\it i.e.}, $f^{+a}_{{\bf s} = a/2}$ describes the irreducible $SU(2)$ multiplet with isospin ${\bf s} = \frac{a}2$.

The natural next issue is to learn how to extract the irreducible $SU(2)$ multiplets from the general reducible ones, e.g.
$\hat{f}^{+a}_{({\bf s} = a/2})$ from $f^{+a}$. A straightforward way is just to  pass to the harmonic expansion
and take into account that every irreducible harmonic monomial is just the eigenfunction of $C$ with the relevant isospin.
One more way to do such a decomposition is to use the harmonic distributions and delta functions, still without applying to the explicit
harmonic expansions. Here we suggest yet a new way of decomposing into the irreducible isospins using merely harmonic derivatives and  not resorting, at any step,
to the explicit harmonic expansions. Just this method can be directly extended to the superspace setting.

For simplicity, we start from the single-charged function $f^+(u)$. Let us define
\beq
\hat{f}^{+}_{(1/2)} = f^+ + {2}\sum_{k=1} (-1)^k \frac{1}{k!(k + 2)!}\,(\partial^{--})^k(\partial^{++})^k\,f^+\,. \lb{A8}
\eeq
Acting on both sides of \eqref{A8} by $\partial^{++}$ and using
\beq
[\partial^{++}, (\partial^{--})^k](\partial^{++})^k\,f^+ = k(k+2)\,(\partial^{--})^{k-1} (\partial^{++})^k\,f^+\,,
\eeq
we find
\bea
\partial^{++}\hat{f}^{+}_{(1/2)} & =& \partial^{++}f^+ + 2\,\sum_{k=1} (-1)^k \frac{1}{(k+1)!(k-1)!}\,(\partial^{--})^{k-1}(\partial^{++})^{k}\,f^+ \nonumber \\
&& + \,2\sum_{k=1} (-1)^k \frac{1}{k!(k + 2)!}\,(\partial^{--})^k(\partial^{++})^{k+1}\,f^+\,.
\eea
The sum of the first two terms cancels the third term, yielding
\beq
\partial^{++}\hat{f}^{+}_{(1/2)} = 0\,\quad \Rightarrow \quad C \,\hat{f}^{+}_{(1/2)} = \frac12\,(\frac12 +1)\hat{f}^{+}_{(1/2)}. \lb{A9}
\eeq
So the ``section'' $\hat{f}^{+}_{(1/2)}$ describes the irreducible isospin ${\bf s} = \frac12$ multiplet. Analogously, we can define
the irreducible parts of $f^+$ carrying higher odd ispospins $3/2, 5/2, \ldots$. This issue will be discussed later on.

Consider now a general function $f^{+ m}$, where $m = 2n$ for integer isospins and $m= 2n+1$ for half-integer isospins. An analog of \eqref{A8} is
\beq
\hat{f}^{+ m}_{(m/2)} = f^{+ m} + \sum_{k=1} (-1)^k \frac{(m + 1)!}{k!(k + m +1)!}\,(\partial^{--})^k(\partial^{++})^k\,f^{+ m}\,. \lb{A86}
\eeq
Once again, it is direct to check that
\beq
\partial^{++} \hat{f}^{+ m}_{(m/2)} = 0\,,\lb{Irred}
\eeq
whence, using the first relation in \eqref{Cas12}, we find
\beq
C \,\hat{f}^{+ m}_{(m/2)} = \frac{m}{2}\,\big(1 +   \frac{m}{2}\big)\,\hat{f}^{+ m}_{(m/2)}\,,
\eeq
thus confirming that $\hat{f}^{+ m}_{(m/2)}$ describe integer fixed isospins $n$ for $m=2n \,(n = 0, 1, 2, \ldots $) and the fixed half-integer ones $n+ \frac12$
for $m= 2n +1\, (n = 0, 1, 2, \ldots)$.

Note that, without loss of generality, in \eqref{A86} we can replace
\bea
f^{+ m} \quad \Rightarrow \quad (\partial^{++})^n \omega  \;\;{\rm for} \;\;m = 2n, \;\; {\rm and} \;\;(\partial^{++})^n \,f^+ \;\;{\rm for} \;\;m = 2n +1.\lb{Repl}
\eea
Using this observation, for writing the whole decomposition of any harmonic function into an infinite sums over the irreducible isospin parts, it suffices to write such sums for $f^+$ and $\omega$.
All others can be recovered by acting on these basic expansions by the proper degrees of $\partial^{++}$. We start with the evident ansatz following from the preservation
of the total harmonic $U(1)$ charges
\beq
\omega = \sum_{k=0} b_k (\partial^{--})^k \hat{f}^{+2k}\,, \qquad f^+ = \sum_{k=0} c_k (\partial^{--})^k \hat{f}^{+(2k+1)}\,, \label{Dec12}
\eeq
where $b_k, c_k$ are some numerical coefficients (undetermined for the moment). To fix these coefficients, we substitute  for $\hat{f}^{+2k}$, $\hat{f}^{+(2k+1)}$ their
expressions from \eqref{A86}, where the replacements
\eqref{Repl} have been performed, and require all terms with derivatives in the right-hand sides of \eqref{Dec12} to be mutually canceled.
Let us briefly indicate the basic steps of this procedure for $f^+$. Considering a few first terms in the decomposition over powers of derivatives, we find that
\beq
c_k = \frac{k+1}{(2k+1)!}\,, \quad  k = 0, 1, 2, \ldots
\eeq
Then we assume that this expression is true for all $C_k$ with $0< k \leq n-1$ and, equating to zero the full coefficient of $(\partial^{--})^n f^+$, find
\beq
c_n = \sum_{k =0}^{n-1} (-1)^k \frac{2(n-k)^2}{(k+1)! (2n + 1-k)!}\,,
\eeq
that can be transformed to
\beq
c_n = -2\sum_{k =1}^{n} (-1)^k \frac{(n-k +1)^2}{l! (2n + 2-k)!}\,. \lb{CnFin}
\eeq
This sum can be evaluated using the identity \cite{Prudnikov}
\beq
\sum_{k =0}^{m} (-1)^k \frac{(m - k)^2 (2m)!}{k! (2m - k)!} = 0\,,\lb{IdentImp}
\eeq
which is valid for any $m\geq 0$. We choose $m=n+1$ and rewrite the latter identity as
\beq
\sum_{k =0}^{n} (-1)^k \frac{(n +1- k)^2 (2n + 2)!}{k! (2n + 2 -k)!} = 0\,.\lb{IdentImp2}
\eeq
Comparing it with \eqref{CnFin}, we finally obtain
\beq
c_n = \frac{n+1}{(2n+1)!}\, \label{coefficients_for_f}
\eeq
for any $n\geq 0$. Analogously, one can prove that
\beq
b
_n = \frac{1}{(2n)!}\,. \label{coefficients_for_omega}
\eeq

The final expressions for the infinite isospin decompositions of $f^+$ and $\omega$ read
\bea
f^+ = \sum_{k=0}\frac{k+1}{(2k+1)!}\,(\partial^{--})^k\,\hat{f}^{+(2k+1)}\,, \qquad \omega  = \sum_{k=0}\frac{1}{(2k)!}\,(\partial^{--})^k\,\hat{f}^{+2k}\,. \lb{fandomDecom}
\eea
Since $\partial^{--}$ commutes with the Casimir operator $C$, the functions appearing in these sums,
\beq
\hat{f}^+_{(k +1/2)} :=(\partial^{--})^k\,\hat{f}^{+(2k+1)}\,, \qquad \hat{\omega}_{(k)} : =(\partial^{--})^k\,\hat{f}^{+2k}\,, \lb{Basfunc}
\eeq
are eigenfunctions of $C$ with the same eigenvalues as for $\hat{f}^{+(2k+1)}$ and $\hat{f}^{+2k}$\,,
\beq
C\,\hat{f}^+_{(k +1/2)} = (k +1/2)(k +3/2) \hat{f}^+_{(k +1/2)}\,, \qquad C\,\hat{\omega}_{(k)}  = k(k+1)\hat{\omega}_{(k)} , \;\;k\geq 0\,,
\eeq
and so correspond to the infinite sets of half-integer or integer isospins ${\bf s} = k +1/2$ and ${\bf s} = k$. Thus eqs. \eqref{fandomDecom} secure
the infinite decompositions of the harmonic functions $f^+$ and $\omega$ over the irreducible $SU(2)$ isospin multiplets. Note that the basic functions, by their definition
\eqref{Basfunc}, satisfy the constraints
\beq
(\partial^{++})^{k+1}\hat{f}^+_{(k +1/2)} =0\,,  \quad (\partial^{++})^{k+1}\hat{\omega}_{(k)} = 0\,.
\eeq

It is easy to find the analogous expansions for general harmonic functions $f^{+m}$ defined in \eqref{Repl}. Acting on the series in \eqref{fandomDecom} with $(\partial^{++})^n$ and using the commutation relations
\eqref{A1}, as well as the irreducibility conditions \eqref{Irred}, we obtain
\be
f^{+m} = \sum_{k=0}\frac{(k+m)!}{k!(2k+m)!}\,(\partial^{--})^k\,\hat{f}^{+(2k+m)}\,. \lb{GenDec+m}
\ee

As the last topic of this Appendix we rewrite the irreducible functions $\hat{f}_{(1/2)}$ and  $\hat{f}_{(m/2)}$ defined in \eqref{A8} and \eqref{A86} directly through the Casimir operator $C$. After some simple algebra
repeatedly using 1st relation in \eqref{Cas12} we find
\beq
\hat{f}^{+}_{(1/2)} = \Big\{1 + {2}\sum_{k=1} (-1)^k \frac{1}{k!(k + 2)!}\,\prod_{l=1}^k\, (C - l^2 + 1/4)\Big\} f^+ \lb{A88}
\eeq
and
\beq
\hat{f}^{+ m}_{(m/2)}
= \Big\{1 + \sum_{k=1} (-1)^k \frac{(m + 1)!}{k!(k + m + 1)!}\,\prod_{l=1}^k\, [C - (l + {m}/2)(l -1 + {m}/2 )]\Big\}f^{+ m}\,. \lb{A89}
\eeq
 All other irreducible pieces of $f^+$ and $\omega$ can be expressed through the
unconstrained functions $f^+$ and $\omega$ (in fact through harmonic derivatives $\partial^{++}$ thereof) in a similar way.  The differential operators appearing in \eqref{A88} and  \eqref{A89} are just the projection
operators on the corresponding irreducible isospins. They can be checked to obey all the standard properties of such operators: square to unity and are orthogonal to the operators extracting other irreducible $SU(2)$
multiplets. Instead of presenting the generic proof, we will show that the operator in \eqref{A88} indeed distinguishes the irreducible isospin $1/2$ component in the infinite sum \eqref{fandomDecom} for $f^+$.

Look first at the isospin $1/2$ part ${f}^+_{(1/2)}$ of $f^{+}$. At any $k$ the products in \eqref{A88}, when acting on ${f}^+_{(1/2)}$, are vanishing due to the vanishing of
the first multiplier, $(C -3/4){f}^+_{(1/2)} = 0$. So, when the whole operator within the curly bracket in \eqref{A88} acts on ${f}^+_{(1/2)}$, it yields just the latter,
and it remains to show that the sum of all products in \eqref{A88}, acting
on any other irreducible term in $f^+$, say, on ${f}^+_{(m -1/2)}, m \geq 2\,,$ yields just $-{f}^+_{(m -1/2)}$ which  cancels ${f}^+_{(m -1/2)}$
in the first term of the curly bracket in \eqref{A88}.

Looking at \eqref{A88} and taking into account that $C\,{f}^+_{(m -1/2)} = (m^2 - 1/4){f}^+_{(m -1/2)}$, we see that in this case the non-zero contributions could come
only from the products
$\prod_{l=1}^k\, (m^2 - l^2), \;k \leq m-1$. So we need to evaluate the sum in
\beq
2\sum_{k=1}^{m-1} (-1)^k \frac{1}{k!(k + 2)!}\,\prod_{l=1}^k\, (m^2 - l^2)\, \hat{f}^+_{(m -1/2)}\,, \quad m \geq 2\,.
\eeq
Using
\beq
\prod_{l=1}^k\, (m^2 - l^2) = \prod_{l=1}^k (m-l)\prod_{l=1}^k (m+l) = \frac{(m-1)!(m+k)!}{(m -k -1)!m!}\,,
\eeq
this sum can be rewritten as
\beq
2\sum_{k=0}^{m-1} (-1)^k \frac{(m+k)! (m-1)!}{k! m!(k + 2)! (m-1 -k)!} - 1 := X -1\,.
\eeq
Changing the summation index as $k \rightarrow m-1-k$, the sum $X$ can be transformed to the expression
\beq
X =  -2(-1)^m \frac{1}{(m+1)m} F, \quad F = \sum_{k=0}^{m-1} (-1)^k \frac{(m+1)!(2m -k -1)!}{k! m!(m+1-k)!(m -1 -k)!}\,.
\eeq
Further, making use of the general identity \cite{Prudnikov},
\beq
\sum_{k=0}^{s}(-1)^k p\frac{(s +p -k-1)!}{k!(p -k)!(s - k)!} = 0\, \quad (s, p\;\; {\rm are \;\;some\;\; integers}),
\eeq
and choosing $p = m+1, s= m-1$, we find
$$
F=0\,\rightarrow \, X=0\,.
$$
Thus indeed the curly bracket in \eqref{A88}, when acting on ${f}^+_{(m -1/2)}$, $m\geq 2$, yields $(1-1){f}^+_{(m -1/2)} =0$. In other words, it acts as the projection operator
extracting the irreducible isospin $\frac12$ part ${f}^{+}_{(1/2)}(u)$ from the unconstrained reducible harmonic function $f^+(u)$. Analogous projection operators can of course
be defined for other irreducible parts of $f^+$ and in the more general case of $f^{+m}(u)$.

Finally, it is worth to point out that in the supersymmetric case there is no such a direct relation between the superisospin and the external harmonic $U(1)$ charge
of the superfield as in the pure harmonic case between the isospin and the harmonic  $U(1)$ charge of functions on $SU(2)/U(1)$. For instance, the harmonic analytic superfields
$V^{++}$ and $\omega$ have the following superisospin contents: $0, 1, 2, ...$ for $V^{++}$ and $1, 2, 3,...$ for $\omega$, which should be compared with the isospin  contents $1, 2, 3, ...$
for $f^{++}$ and $0, 1, 2, 3...$ for $f^0 = \omega$.

\section{Projectors via infinite products}

In Appendix A we have shown how to expand an arbitrary function $f^{+q}(u)$ defined on the coset $SU(2)/U(1)$ in an infinite sum of irreducible isospins represented
by the properly constrained harmonic functions. In particular, for $q=1$,
\begin{equation}
    f^{+} = {f}^{+}_{(1/2)} + {f}^{+}_{(3/2)} + {f}^{+}_{(5/2)} + ...\,,
\end{equation}
\bea
{f}^{+}_{(1/2+m)} = \frac{(m+1)}{(2m+1)!}\sum_{k=0}^{\infty} \frac{(-1)^k(2m+2)!}{k!(k+2m+2)!}
(\partial^{--})^{k+m}(\partial^{++})^{k+m} f^{+}, \qquad m \in \{0, 1, 2, ...\}.\label{Ab41}
\eea

On the other hand, recall the general formula for projector operators on the fixed eigenvalue $\lambda_{i}$ of some operator $C$:
\begin{equation}\label{Deriv/Proj/ExpressionProjector}
    \text{П}_{(i)} = \prod_{k\ne i} \frac{C-\lambda_k}{\lambda_i - \lambda_k}.
\end{equation}
In our case we deal with the irreducible isospin components of $f^{+}$, which obey the eigenvalue equations $C\hat{f}^{+}_{(\bf j)} = {\bf j}({\bf j} +1)\hat{f}^{+}_{(\bf j)}$.
So the general formula (\ref{Deriv/Proj/ExpressionProjector}) amounts to:
\begin{equation}
    \hat{f}^{+}_{(1/2+m)} = \text{П}_{({\bf s}=1/2+m)}^{(a=1)} f^{+}, \quad \text{П}_{({\bf s}=1/2+m)}^{(a=1)} = \prod_{j\ne {\bf s}}
    \frac{C - {\bf j}({\bf j}+1)}{{\bf s}({\bf s}+1) - {\bf j}({\bf j}+1)}, \quad {\bf j} \in \{1/2,\ 3/2,\ ...\}\,. \label{Ab42}
\end{equation}
The natural question is as to whether the projection operator in (\ref{Ab41}) coincides with that defined in (\ref{Ab42}).
In this Appendix we present the explicit connection between the expressions (\ref{Ab41}) and (\ref{Ab42})
for the simplest case of $\hat{f}^{+}_{(1/2)}$ with the isospin ${\bf s}=1/2$.

Our starting point will be the expression (\ref{Ab42}) for the particular choice ${\bf j} = n - 1/2$ and ${\bf s}=1/2$:
\begin{equation}
\text{П}_{(1/2)}^{(a=1)} = \prod_{n=2}^{\infty}\frac{C-\left(n^2 - \frac{1}{4}\right)}{\frac{3}{4}-
\left(n^2 - \frac{1}{4}\right)} = \prod_{n=2}^{\infty}\frac{C-\left(n^2 - \frac{1}{4}\right)}{1 - n^2}\,. \label{Aa1}
\end{equation}
Now, consider the finite part of (\ref{Aa1}) with $N$ factors, which we label as $\left(\tilde{\text{П}}_{(1/2)}^{(a=1)}\right)^{(N)}$:
\begin{align}
\left(\tilde{\text{П}}_{(1/2)}^{(a=1)}\right)^{(N)} &= \prod^{N+1}_{n=2}\frac{C-(n^2-1/4)}{1-n^2} \nonumber \\
&= \prod^{N+1}_{n=2}\frac{(-1)}{n^2-1}\prod^{N+1}_{n=2} {C-(n^2-1/4)} \nonumber \\
&= \frac{2!(-1)^N}{N! (N+2)!}\prod^{N+1}_{n=2} {C-(n^2-1/4)}. \label{I,3}
\end{align}
Notice that the expression (\ref{I,3}) closely resembles the second  item in (\ref{A88}). Let us build on this.
Albeit we deal with  a finite \textit{product} \eqref{Aa1}, we can equally represent it as a finite \textit{sum}. Assume that this partial product
amounts to some sum $S^{(N)}$ (other indices are hidden for a moment):
\begin{equation}\label{product_sum_relation}
    \text{П}^{(N)} = S^{(N)}, \quad S^{(N)} = \sum_{k=0}^{N}A^{(k)}.
\end{equation}
One can recover each term from this sum by the simple formula:
\begin{equation}
    A^{(k)} = S^{(k)} - S^{(k-1)} = \text{П}^{(k)} - \text{П}^{(k-1)}, \quad A^{(0)}= 1\,.
\end{equation}
Substituting the expressions (\ref{I,3}) for the products in this formula, we derive:
\begin{align}
    A^{(k)} &= \frac{2!(-1)^k}{k! (k+2)!}\left[\prod^{k}_{n=2} {C-(n^2-1/4)}\right]\left[C-(k^2-1/4) + (k-1)(k+1)\right] \\
    &= \frac{2!(-1)^k}{k! (k+2)!}\left[\prod^{k}_{n=1} {C-(n^2-1/4)}\right]. \label{miracle}
\end{align}
Then we collect all terms $A^{(k)}$ into the sum (\ref{product_sum_relation}):
\begin{equation}
    \text{П}_{(1/2)}^{(a=1)} \overset{(\ref{product_sum_relation})}{=}1 + \sum_{k=1}^{\infty}\frac{2!(-1)^k}{k!(k+2)!}\prod_{l=1}^{k}(C-l^2+1/4),
\end{equation}
and compare it with (\ref{A88}). Both expressions coincide, though were derived in different ways.

The same logic applies to other values of isospins ${\bf s}$, as well as the charges $+a$.
Here we present only the final results. For even charges:
\begin{align}
\boxed{\text{П}^{(a=2m)}_{({\bf s}=k)} = \frac{(k+m)!}{(k-m)!}
\sum_{n=0}^{\infty}\frac{(-1)^n (2k+1)}{n!(n+2k+1)!}(\partial^{--})^{n+k-m}(\partial^{++})^{n+k-m}, \quad k\ge m\,.} \label{product_deriv_15}
\end{align}
Analogously, for odd charges:
\begin{align}
\boxed{\text{П}^{(a=2m-1)}_{({\bf s}=k-1/2)} = \frac{k(k+m-1)!}{(k-m)!}
\sum_{n=0}^{\infty}\frac{2(-1)^n}{n!(n+2k)!}(\partial^{--})^{n+k-m}(\partial^{++})^{n+k-m}, \quad k\ge m >0\,.} \label{product_deriv_17}
\end{align}
After proper relabeling, one recovers (\ref{Ab41}), as well as \eqref{A88}, \eqref{A89} and \eqref{A86}. For both sums \eqref{product_deriv_15} and \eqref{product_deriv_17} one can find
alternative representations through infinite products following the procedure outlined above.

Note that in the process of calculations we used the following general identity:
\begin{equation}
    (\partial^{++})^{n}(\partial^{--})^{n} f^{+a} = \prod_{k=1}^{n} \Big[ C - \left(k-1-\frac{a}{2} \right)\left(k-\frac{a}{2}\right)\Big]f^{+a}\,.
\end{equation}

\end{document}